# A Parametric and Feasibility Study for Data Sampling of the Dynamic Mode Decomposition— Range, Resolution, and Universal Convergence States

Cruz Y. Li[1] (李雨桐), Zengshun Chen[2*](陈增顺), Tim K.T. Tse[3**](谢锦添), Asiri Umenga Weerasuriya[4], Xuelin Zhang[5] (张雪琳), Yunfei Fu[6](付云飞), Xisheng Lin[7] (蔺习升)

[1,3,4,6,7] *Department of Civil and Environmental Engineering, The Hong Kong University of Science and Technology, Hong Kong SAR, China*

[2] *Department of Civil Engineering, Chongqing University, Chongqing, China*

[5] *School of Atmospheric Sciences, Sun Yat-sen University, Zhuhai, China.*

[1] yliht@connect.ust.hk; ORCID 0000-0002-9527-4674
[2] zchenba@connect.ust.hk; ORCID 0000-0001-5916-1165
[3] timkttse@ust.hk; ORCID 0000-0002-9678-1037
[4] asiriuw@connect.ust.hk; ORCID 0000-0001-8543-5449
[5] zhangxlin25@mail.sysu.edu.cn; ORCID 0000-0003-3941-4596
[6] yfuar@connect.ust.hk; ORCID 0000-0003-4225-081X
[7] xlinbl@connect.ust.hk; ORCID 0000-0002-1644-8796

[*] Co-first author with equal contribution.
[**] Corresponding author
All correspondence is directed to Dr. Tim K.T. Tse.

# Abstract

Scientific research and engineering practice often require the modeling and decomposition of nonlinear systems. The Dynamic Mode Decomposition (DMD) is a novel Koopman-based technique that effectively dissects high-dimensional nonlinear systems into periodically distinct constituents on reduced-order subspaces. As a novel mathematical hatchling, the DMD bears vast potentials yet an equal degree of unknown. This effort investigates the nuances of DMD sampling with an engineering-oriented emphasis. It aimed at elucidating how sampling range and resolution affect the convergence of DMD modes. We employed the most classical nonlinear system in fluid mechanics as the test subject—the turbulent free-shear flow over a prism—for optimal pertinency. We numerically simulated the flow by the dynamic-stress Large-Eddies Simulation with Near-Wall Resolution. With the large-quantity, high-fidelity data, we parametrized and identified four global convergence states: *Initialization, Transition, Stabilization*, and *Divergence* with increasing sampling range. Results showed that *Stabilization* is the optimal state for modal convergence, in which DMD output becomes independent of the sampling range. The *Initialization* state also yields sufficient accuracy for most system reconstruction tasks. Moreover, defying popular beliefs, over-sampling causes algorithmic instability: as the temporal dimension, $n$, approaches and transcends the spatial dimension, $m$ (i.e., $m < n$), the output diverges and becomes meaningless. Additionally, the convergence of the sampling resolution depends on the mode-specific dynamics, such that the resolution of 15 frames per cycle for target activities is suggested for most engineering implementations. Finally, a bi-parametric study revealed that the convergence of the sampling range and resolution are mutually independent.

# Keywords

# Declarations

# Nomenclature

The list contains the nomenclature employed in this study.

*Latin Letters*

| | |
|---:|---|
| $A$ | Infinite-dimensional Koopman operator |
| $\tilde{A}$ | Similarity-matrix approximate of $A$ |
| $a_e$ | $a_e$=0.53. Empirical constant for $LES_{IQ}$ |
| $\alpha_v$ | $a_v$=0.05. Empirical constant for $LES_{IQ}$ |
| $b$ | Scalar Magnitude of Z-transform $z=be^{j\theta}$ |
| $C_{D, RMS}$ | Root-mean-square deviation of Drag coefficient |
| $\langle C_D \rangle$ | Mean Drag coefficient |
| $C_L$ | Lift coefficient |
| $C_{L, RMS}$ | Root-mean-square deviation of Lift coefficient |
| $C_L{}'$ | Fluctuating component of Lift coefficient |
| $C_p$ | Wall pressure coefficient |
| $C_s$ | Smagorinsky constant |
| $\mathcal{C}_s$ | Dynamic Smagorinsky coefficient |
| $D$ | Side length of prism/diameter of cylinder |
| $\boldsymbol{D}_\alpha$ | Diagonal matrix of $\boldsymbol{\alpha}$ |
| $d$ | Distance from a cell centroid to the closest wall |
| $E_{CL}{}'$ | Spectrum of fluctuating Lift coefficient |
| $E_R$ | Portion of $E_{TKE}$ resolved by the filtered Navier-Stokes Equations |
| $E_{TKE}$ | Spectrum of total turbulence kinetic energy |
| $\|e\|_{2, ins}$ | Mean $l_2$-norm of reconstruction error by DMD |
| $G$ | Hexahedral grid of 4.3 million elements |
| $G(x)$ | Filter function in three-dimensional space |
| $G_{\|e\|_2}$ | Grand mean $l_2$-norm of reconstruction error by DMD |
| f.p.c. | Frames per cycle |
| $g_j$ | Growth rate of DMD mode $\boldsymbol{\phi}_j$ |
| $I_j$ | I-criterion for dominant mode selection |
| $\Im$ | The imaginary space |
| $k_{all}$ | Total turbulence kinetic energy |
| $k_{num}$ | Pseudo-energy term for discretization error/numerical residual |
| $k_r$ | Portion of $k_{all}$ resolved by the filtered Navier-Stokes Equations |
| $k_{sgs}$ | Portion of $k_{all}$ modelled by subgrid-scale model(s) |
| $LES_{IQ}$ | Index quantifying the resolution of a LES grid |
| $\mathcal{L}_{ij}$ | The Germano identity relating grid- and test-filtered stresses |
| $\mathcal{L}_{ij}^d$ | Deviatoric component of $\mathcal{L}_{ij}$ |
| $\mathcal{L}_{ij}^{S}$ | Smagorinsky model for $\mathcal{L}_{ij}^{d}$ |
| $l_s$ | Smagorinsky lengthscale |
| $l_x$ | Characteristic length of a boundary layer |
| $M_{ij}$ | A definition for least square analysis after Lilly [1] |
| $m$ | DMD temporal dimension (*i.e.*, number of snapshots) |
| $N_G$ | Spatial dimension of $G$ |
| $N_R$ | Spatial dimension of $R$ |
| $n$ | DMD spatial dimension (*i.e.*, number of nodes per snapshot) |
| $\bar{p}$ | Modified pressure |
| $R$ | Region with refined hexahedral grid within $G$ |
| $r$ | Order of truncation |
| $Re$ | Reynolds number based on $D$ |
| $Re_x$ | Reynolds number based on $l_x$ |
| $\Re$ | The real space |
| $St\ (St_j)$ | ($j^{th}$) Strouhal number. Reduced frequency. |
| $\overline{S}_{ij}$ | Filtered rate of strain |
| $\overline{S}$ | Filtered characteristic rate of strain |
| $\mathcal{T}_{ij}^{SGS}$ | Test-filtered stress |
| $\mathcal{T}_{ij}^{sgs}$ | Deviatoric component of $\mathcal{T}_{ij}^{SGS}$ |
| $t*$ | Normalized time-step for DMD sampling |
| $t_{DNS}$ | Time step of Detached-Eddy Simulation |
| $t_{DNS}$ | Time step of Direct Numerical Simulation |
| $t_{washout}$ | Time step of washout |
| $\boldsymbol{U}$ | Matrix containing all POD modes |
| $\boldsymbol{u}$ | Instantaneous velocity field |
| $u$ | Flow velocity in the x-direction |
| $\langle U \rangle$ | Mean magnitude of flow velocity |
| $\langle \boldsymbol{u} \rangle$ | Mean component of $\boldsymbol{u}$ |
| $U(z)$ | Z-transform of $u[k]$ |
| $U_\infty$ | Free-stream flow velocity |
| $\boldsymbol{u_j}$ | Component of $\boldsymbol{U}$, $j^{th}$ POD mode |
| $u[k]$ | Arbitrary discrete signal as input of Z-transform |
| $\boldsymbol{u'}$ | Fluctuating component of $\boldsymbol{u}$ |
| $u^{'}$ | Fluctuating component of $u$ |
| $\overline{u}_i$ | Filtered velocity |
| $V$ | Volume of computational cell |
| $\boldsymbol{V}$ | Matrix containing temporal information of spatial matrix $\boldsymbol{U}$ |
| $\boldsymbol{V_{and}}$ | The Vandermonde matrix |
| $v$ | Flow velocity in the y-direction |
| $\boldsymbol{v_j}$ | Component of $\boldsymbol{V}$, temporal evolution of $\boldsymbol{u_j}$ |
| $v^{'}$ | Fluctuating component of $v$ |
| $\boldsymbol{W}$ | Matrix containing all eigenvectors of $\tilde{\boldsymbol{A}}$ |
| $w$ | Flow velocity in the z-direction |
| $w_j$ | Component of $W$, $j^{th}$ eigenvector of $\tilde{A}$ |
| $w^{'}$ | Fluctuating component of $w$ |
| $X_1$ | Input snapshot sequence spanning from $1$ to $m\text{-}1$ |
| $X_2$ | Input snapshot sequence spanning from $2$ to $m$ |
| $x_{DMD,k,i}$ | DMD reconstructed data at node $k$ and instant $i$ |
| $x_i$ | Component of $X_1$ and $X_2$, individual input snapshot |
| $x_{k,i}$ | Original input data at node $k$ and instant $i$ |
| $x^+$ | Non-dimensional wall distance in the x-direction |
| $y^+$ | Non-dimensional wall distance in the y-direction |
| $z^+$ | Non-dimensional wall distance in the z-direction |
| $\mathcal{Z}$ | Z-transformation |

*Greek Letters*

| | |
|---|---|
| $\boldsymbol{\alpha}$ | Modal amplitude, $\alpha$-criterion for dominant mode selection |
| $\alpha_j$ | Modal amplitude of DMD mode $\boldsymbol{\phi}_j$ |
| $\Delta$ | Grid-dependent filter of LES |
| $\Delta t$ | Time-step of LES simulation in second |
| $\Delta t^*$ | Normalized time-step of LES simulation |
| $\Delta x_{min}$ | Minimum x-dimension of wall-adjacent cells |
| $\Delta y_{min}$ | Minimum y-dimension of wall-adjacent cells |
| $\Delta z$ | z-dimension of cells |
| $\widetilde{\Delta}$ | Secondary test filter of dynamic stress model |
| $\delta_{BL,99\%}$ | Laminar boundary layer thickness at 99% of $U_\infty$ |
| $\delta_{ij}$ | The Kronecker Delta |
| $\kappa$ | The von Kármán constant $\kappa = 0.40$ |
| $\Lambda$ | Matrix containing all eigenvalues of $\tilde{A}$ |
| $\lambda_j$ | Component of $\Lambda$, $j^{th}$ eigenvalue of $\tilde{A}$ |
| $v$ | Fluid molecular kinematic viscosity |
| $v_{sgs}$ | Subgrid viscosity |
| $\rho$ | Fluid density |
| $\boldsymbol{\Sigma}$ | Diagonal matrix containing all singular values of matrix $U$ |
| $\sigma_j$ | Component of $\boldsymbol{\Sigma}$, singular value (energy content) of $\boldsymbol{u}_j$ |
| $\tau_{kk}^{SGS}$ | Isotropic component of $\tau_{ij}^{SGS}$ |
| $\tau_{ij}^{SGS}$ | Subgrid stress |
| $\tau_{ij}^{sgs}$ | Deviatoric component of $\tau_{ij}^{SGS}$ |
| $\widetilde{\tau_{ij}^{SGS}}$ | Test-filtered subgrid stress |
| $\boldsymbol{\Phi}$ | Matrix containing all DMD modes |
| $\boldsymbol{\phi}_j$ | Component of $\boldsymbol{\Phi}$, $j^{th}$ DMD mode (complex shape) |
| $\omega_j$ | Continuous-time frequency of DMD mode $\boldsymbol{\phi}_j$ |

*Others*

| | |
|---|---|
| $^T$ | Conjugate transposition |
| † | Moore-Penrose pseudoinverse |
| $\Vert \ \Vert_F$ | Frobenius Normalization |
| $SF$ | Binary index of sampling resolution |
| **Bold** | Vector quantity |
| Normal | Scalar quantity |
| $i,j,k,l$ | Indicial notation for vector/tensor quantities |

# 1. Introduction

In today's realm of science and engineering, nonlinearity remains one of the few unanswered questions of classical physics. Nonlinear systems are often high-dimensional and have intertwining dynamics, so modeling them can be extremely strenuous. Seeking engineering solutions, applied mathematicians invented the Reduced-Order Models (ROM) that serve precisely for the purpose of dimension reduction and modeling of nonlinear systems. After decades, ROMs proved effective and crucial for nonlinear analysis. In fact, its value in data processing is so prominent that it has already become a well-defined discipline with a vast collection of literature [2].

The Dynamic Mode Decomposition (DMD) is a new addition to the ROM family [3], [4]. Like its closest cousin, the Proper Orthogonal Decomposition (POD) [5], the DMD is purely data-driven. This undiscriminating nature permits broad applications. Kutz *et al.* [6] offered an excellent collage of DMD implementations on nonlinear systems from fluid mechanics, video processing, signal and controls, epidemiology, neuroscience, finance, *etc.* For most systems, the DMD proved powerful in generating spatiotemporally accurate representations of complex dynamics and, subsequently, in producing visual decompositions with insightful revelations [4], [7]–[11]. Nonetheless, like any other brand-new mathematical hatchlings, its potential comes with many unknowns.

Perhaps the greatest uncertainty that prohibits the pervasive dissemination of the DMD in engineering practices is the nuances in data sampling. Till today, most contributions to the DMD literature come directly from applied mathematicians. Their expertise in linear algebra and signal processing translates to intuitive decisions on the sampling state space, range, resolution, truncation order, *etc.* However, it is unrealistic to expect these mathematical intuitions from most engineering practices. Moreover, the transition from a priori findings to practical successes may not always be straightforward. For example, Klus *et al.* [12] and Korda and Mezić [13] laid the theoretical basis for the sampling convergence on the observable space and state space, respectively, but the realization of such convergence in practice remains a knowledge gap.

In this effort, we look to bridge some major gaps in the DMD sampling. Specifically, we offer a parametric, *a posteriori* investigation on the sampling range and resolution. To this end, Schmid briefly discussed the sampling convergence as the dimension of the Krylov subspace in the DMD's debut [4]. However, the shortage of a dedicated follow-up led to vast uncertainties in the subsequent literature. Take the fluid mechanics community as an example, there are major inconsistencies in the data sampling of DMD implementations (**Table** 1). The sampling range varies from as few as *20* snapshots to as large as *89* oscillation cycles. The sampling resolution, on the other hand, lacks any discernible uniformity, accruing in units from the non-dimensional timescale frame-per-cycle (f.p.c) to the numerical timescale $t_{DNS}$ to the physical timescale Hertz. Some did not even specify a clear resolution scale. In terms of convergence, although some studies relied solely on the $l_2$-residual for assessment, most works neglected the convergence assessment altogether. The inconsistencies in data sampling, as observed by our parametric study and presented subsequently, resulted in substantial differences in DMD output. Therefore, this alarming gap needs to be filled with precise knowledge of sampling range, resolution, and convergence.

**Table 1** Examples of DMD implementations on fluid systems with various sampling range and resolution.

| Sampling Range [snapshots] | Sampling Resolution* | Configuration | Contribution |
|---|---|---|---|
| 251 | 2 $t_{DNS}$ | Jet flow | Rowley *et al.* [3] |
| 20-30 | u.c. | Prism wake | Schmid *et al.* [4] |
| 500 | 2000 Hz | Flexible membrane wake | |
| 280 | 280 f.p.c. | Cylinder wake | Jardin and Bury [14] |
| 3000 | 20 $t_{DES}$ | Wall-mounted cube | Muld *et al.* [15] |
| 89 cycles | 112 f.p.c. | Cylinder wake | He *et al.* [16] |
| 1000 | 1000 Hz | Cylinder wake | Tissot *et al.* [17] |
| 896 | 40 f.p.c. | Wind turbine | Sarmast *et al.* [18] |
| 1200 | 45 $t_{washout}$ | Pipe flow | Gómez *et al.* [19] |

| 501 | u.c. | Transitional Jet | Roy *et al.* [20] |
| 123 | 30 Hz | Bluff-body wakes | Wan *et al.* [21] |
| 300 | 17 f.p.c. | SD7003 airfoil | Ducoin *et al.* [22] |
| 300 | u.c. | Transonic backward step | Statnikov *et al.* [23] |

*u.c. – unclear; f.p.c – frames per cycle

Intending to fill this gap, we investigated the sampling convergence by significantly extending the sampling range from $2.0 \times 10^3$ to as large as $2.3 \times 10^4$ snapshots and refining the resolution from 50 to as fine as $10^5$ Hz. We also inspected both the global and mode-specific behaviors of the DMD algorithm for the convergence assessment. Importantly, we conducted our investigation using the canonical turbulent free-shear flow for the optimal reproducibility and pertinency to other nonlinear systems. We employed a moderately-high subcritical flow regime at $Re=2.2 \times 10^4$ in a prism wake for the parametric study herein.

We chose the bluff-body wake as the test subject with several justifications. As free-shear flows (*i.e.,* the velocity of one direction dominates), the conclusions shall apply to a wide range of flow scenarios, such as jets or mixing layers, as they belong to the same class and are behaviorally similar. By including the upstream bluff body, the applicability further extends to many wall-bounded boundary layer flows. Moreover, the prism wake is simplistic yet sophisticated enough for a parametric study. The simplicity refers to the configuration's commonality in engineering application and the abundance of knowledge we share on its fluid mechanism [24]–[28]. The sophistication refers to its simultaneous inclusion of many fluid phenomena like stagnation, separation, reattachment, secondary separation, vortex roll-up, *etc*.

On the other hand, we also selected a moderately high Reynolds number. This subcritical regime is during the shear layer transition and remains phenomenologically similar for a significant range of *Re* [29]. The selected *Re* also accommodates high-fidelity simulation with common computing power for engineering applications. To this end, we numerically simulated the flows by the Large-Eddy Simulations with Near-Wall Resolution (LES-NWR) [30]. Rigorous tests on the numerical results certify the fidelity of the DMD input.

This comprehensive parametric effort expended tremendous computational resources. On two 64- and 80-core high performance computers (HPCs), the investigation took over *16* months to complete, tallying over *800,000* core-hours excluding those dedicated to the LES-NWR, and more than *1.5* million core-hours altogether. Investing the vast resource, we aim to provide an insightful and reliable reference for future applications of the DMD, especially with its rapid dissemination in the engineering domain.

In composition, we lay out the contextual introduction in Section 1, formulate the DMD algorithm in Section 2, describe and validate our LES-NWR simulation in Section 3, establish a benchmark on the sampling range in Section 4, examine the effect of the sampling range with a bi-parametric study in Section 5, and finally conclude the major findings in Section 6.

## 2. Dynamic Mode Decomposition

We introduce the conceptual and mathematical formulations of the vanilla Dynamic Mode Decomposition in this section. Conceptually, one may reckon the DMD as a joint Koopman-modal analysis. In 1932, Bernard O. Koopman [31] outlined the possibility to represent a nonlinear dynamical system in terms of an infinite-dimensional linear operator acting on a Hilbert space of measurement functions of the state of the system [32]. The Koopman Operator Theory later developed into a whole mathematical subdiscipline for nonlinear systems. To this end, by applying the purely data-driven, physics-uninformed DMD technique, the Koopman analysis linearly approximates the dynamics that evolve the input data in time, and the modal analysis decomposes the entwined Koopman operator into discernible features for interpretation. For simplicity, we dissect the conceptual process of the DMD into five parts, as illustrated by **Fig.** 1.

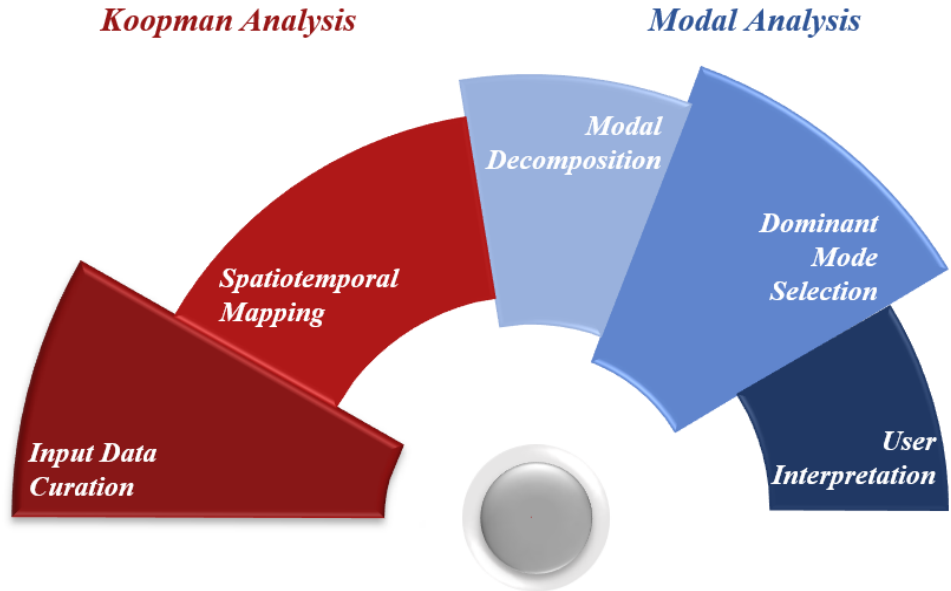

**Fig. 1** An illustration of the conceptual process of the Dynamic Mode Decomposition

## 2.1 Input Data Curation

The first step is the curation of the input data. The DMD's data-driven nature appeals to neural networks in machine-learning. Therefore, the attribute of the input signal is essentially unbounded. For the vanilla DMD, the input signal shall be 1) of transient nature, 2) sampled by a uniform frequency, and 3) captured on a fixed domain. Inadequate input may result in numerical degeneration. Researchers have recently developed some DMD variants to target these input constraints, though by different degrees of success and practicality. We refer readers to Schmid [4] and Kutz *et al.* [6] for non-uniform sampling and Erichson *et al.* [33] for unfixed signal domains.

The input signal may also require noise elimination. The DMD is highly sensitive to frequency content, so excessive contamination may lead to the mistreatment of noise as the dynamics of high-wavenumber turbulence. Thus, noise elimination is critical for experimental data. By contrast, while numerical error (due to discretization and residual) is possible in CFD data, noise is unlikely. With proper numerical settings, the DMD algorithm can also sieve out the erroneous dynamics in the order-reduction and the dominant mode selection processes. For these reasons, we justify our preference of numerical data for the present work. We also refer readers to Dawson *et al.* [34] and Hemati *et al.* [35] for noise elimination techniques in the DMD.

With properly treated data, one shall arrange the input signal into two snapshot sequences, such that:

$$X_1 = \{x_1, x_2, x_3, ..., x_{m-1}\}$$

*(2.1.1)*

$$X_2 = \{x_2, x_3, x_4, ..., x_m\}$$

*(2.1.2)*

where $x_i \in \mathbb{C}^n$ are snapshots sampled at a uniform interval $t^*$. The spatial dimension $n$ corresponds to that of the fixed sampling domain (*i.e.,* number of data entries per snapshot). The temporal dimension $m$ corresponds to that of the time series (*i.e.,* number of snapshots).

## *2.2 Spatiotemporal Mapping*

Spatiotemporal mapping is the core of the Koopman analysis. We suppose a mapping matrix, or a Koopman operator, $A$, that connects the matrices $X_1$ and $X_2$:

$$X_2 = AX_1$$

*(2.2.1)*

$A$ encompasses all the dynamics to evolve a nonlinear system by a single time step. In the scenario herein, $A$ is closely an implicit representation of the Navier-Stokes equations.

We use the adverb 'closely' because, as **Eq.** 2.2.1 implies, $A$ is strictly linear. It is to say, while $A$ is exact for a linear system, it is only a best-fit approximate for a nonlinear system. Even so, it is of tremendous merit. As the readers may be aware, acquiring exact solutions or representations for many nonlinear systems is often a herculean, if not impossible, task. The yet-to-be claimed Millenium Prize for the Navier-Stokes equations is a prime example of such a conundrum. Although $A$ is linear, it becomes an increasingly more accurate representation of a nonlinear system with increasing dimensionality and resolution. In an analogous sense, the Koopman approximation is a numerical discretization of the nonlinear dynamics. In the absence of an analytical solution, the numerical solution provides the most convincing alternative, especially for engineering practices.

One shall also reckon the acquisition of $A$ is by no means straightforward. The elegance of the DMD exudes precisely from the mathematical possibility that it offers in this aspect [20], [21], [25], [26]. Compared to the companion matrix formulation [4], the approximation of $A$ using a similarity matrix [6], [36] proves more robust and tractable for high-dimensional systems [37]. Accordingly, one may assume a low-dimensional similarity matrix $\tilde{A}$ for the approximation of the full-rank $A$.

To obtain $\tilde{A}$, one first performs the Singular Value Decomposition (SVD), economy-sized to avoid null spaces, on $X_1$:

$$X_1 = U\Sigma V^T$$

$$(2.2.2)$$

where $U \in \mathbb{C}^{n \times r}$ contains spatially orthogonal modes $u_j$ on an optimal POD subspace; $\Sigma \in \mathbb{C}^{r \times r}$, a diagonal matrix, contains singular values $\sigma_j$ that describe the modal energy of $u_j$; $V \in \mathbb{C}^{m \times r}$ contains temporally orthogonal modes $v_j$ which pertains to the evolution of $u_j$; the superscript $^T$ denotes the conjugate transposition; $r$ denotes the truncation rank. One may also manipulate the rank of $\tilde{A}$, hence the dimensionality of the low-order approximate, by adjusting $r$.

With it, the POD-projected $\tilde{A} \in \mathbb{C}^{r \times r}$ relates to $A$ by:

$$A = U\tilde{A}U^T$$

$$(2.2.3)$$

Recall **Eq.** 2.2.1, it is intuitive to minimize the difference between $X_2$ and $AX_1$ to find the optimal subspace:

$$\underset{A}{minimize} \ \|X_2 - AX_1\|_F^2$$

$$(2.2.4)$$

where $\| \ \|_F$ denotes the Frobenius normal.

One may re-express **Eq.** 2.2.4 by substituting in **Eqs.** 2.2.2 and 2.2.3:

$$\underset{\tilde{A}}{minimize} \ \left\|X_2 - U\tilde{A}\Sigma V^T\right\|_F^2$$

$$(2.2.5)$$

Finally, one acquires $\tilde{A}$ based on the approximate:

$$A \approx \tilde{A} = U^T X_2 V \Sigma^{-1}$$

$$(2.2.6)$$

And the reduced-order similarity matrix $\tilde{A}$ replaces $A$ in **Eq.** 2.2.1:

$$X_2 = \widetilde{A} X_1$$

**Eq.** 2.2.7 provides an all-inclusive, finite-dimensional, linearly approximated, and highly accurate representation for the dynamics of a nonlinear system.

## 2.3 Modal Decomposition

The acquisition of $\widetilde{A}$ signals one's full possession of a system's spatiotemporal information. The modal decomposition aims to dissect the implicit information into discernible constituents, thus permitting user interpretations.

A commonplace technique to analyze a dynamical model $\widetilde{A}$ is the eigen decomposition:

$$\widetilde{A} W = W \Lambda$$

(2.3.1)

where $W$ contains the eigenvectors $w_j$, and $\Lambda$ contains the corresponding discrete-time eigenvalues $\lambda_j$.

The eigen tuples yield the *exact* DMD modes [36]:

$$\Phi = X_2 V \Sigma^{-1} W$$

(2.3.2)

where $\Phi$ contains the mode shape $\phi_j$.

Every mode $\phi_j$ corresponds to a physical frequency $\omega_j$ in continuous time:

$$\omega_j = \Im\left\{log(\lambda_j)\right\}/t^*$$

(2.3.3)

and a modal growth rate $g_j$:

$$g_j = \Re\left\{log(\lambda_j)\right\}/t^*$$

(2.3.4)

## 2.4 Dominant Mode Selection

The singular value $\sigma_j$, while dictating the modal dominance of POD modes, merely offers a reference to the DMD truncation. The modal dominance of the DMD involves more complications because DMD modes are only on orthogonal temporally and non-orthogonal spatially. The original DMD modal amplitude [36] or the $\alpha$-criterion [8], [9] projects DMD modes onto the input signal, mapping out weighted coefficients $\boldsymbol{\alpha}$ based on the initial conditions:

$$\boldsymbol{\alpha} = \boldsymbol{\Phi}^{\dagger} \boldsymbol{x_1}$$

$(2.4.1)$

where the superscript $^{\dagger}$ denotes the Moore-Penrose pseudoinverse.

The amplitude $\alpha_j$ dictates the modal dominance of DMD modes. However, coefficients based on initial conditions can be obtuse for highly capricious systems, especially those involving random transience. Therefore, Kou and Zhang [37] proposed an evolution-informed amplitude, or the $I$-criterion, for modal dominance. One shall obtain the Vandermonde matrix $\boldsymbol{V_{and}}$, such that:

$$\boldsymbol{X_1} = \boldsymbol{\Phi} \boldsymbol{D_\alpha} \boldsymbol{V_{and}} = [\boldsymbol{\phi_1}, \ \boldsymbol{\phi_2}, \ ..., \ \boldsymbol{\phi_r}] \begin{bmatrix} \alpha_1 & & & \\ & \alpha_2 & & \\ & & \ddots & \\ & & & \alpha_r \end{bmatrix} \begin{bmatrix} 1 & \lambda_1 & \cdots & \lambda_1^{i-1} \\ 1 & \lambda_2 & \cdots & \lambda_2^{i-1} \\ 1 & \vdots & \ddots & \vdots \\ 1 & \lambda_r & \cdots & \lambda_r^{i-1} \end{bmatrix}$$

$(2.4.2)$

where $i$ denotes time instant in the input signal.

One then obtains the $I$-amplitude $I_j$ by:

$$I_j = \sum_{i=1}^{N} \left| \alpha_j \ \lambda_j^{i-1} \right| \ \left\| \boldsymbol{\phi_j} \right\|_F^2 \ \varDelta t$$

$(2.4.3)$

We remind readers that in addition to those introduced herein, an array of algebraic criteria or deep-learning techniques, each with advantages and limitations, have been developed for the selection of the most spatiotemporally dominant DMD modes [35], [38]–[40]. Users may employ various criteria accordingly to preference and analytical need.

## 2.5 User Interpretation

Being physics-uninformed, the output of the DMD demands user interpretation. Needless to reiterate the importance of acumen and background knowledge, we briefly comment on the mathematical implications of the DMD in hope to inspire better insights into its essence.

In the humblest scenario, suppose we have a fluid system in the steady state. The sampling of the flow field snapshots inevitably yields discrete signals. Meanwhile, by finding a linear approximation to the nonlinear fluid dynamics, the DMD produces a linearly time-invariant (LTI) system. Therefore, in the eigen space, the seek for complex frequencies is equivalent to finding eigenfunctions of a linear operator (*i.e.*, $A$ or $\tilde{A}$). Similarly, DMD modes render poles on a Region of Convergence (ROC) of a Z-transformation [2], where if we suppose an input data $u[k]$, then $U(z)$ becomes:

$$U(z) = \mathcal{Z}\{u[k]\} = \sum_{k=0}^{\infty} u[k]z^{-k}$$

*(2.5.1)*

where $k \in \mathbb{Z}^+$ yields a unilateral Z-transform and $z = be^{j\theta}$.

As such, the DMD modes, or the poles, contain vast information about the LTI system like stability, causality, *etc.* The DMD also intakes data of intertwined dynamics, and sorts them out in terms of frequency-content, and returns distinct representations of periodicities in the form of Ritz pairs. Therefore, for the turbulence system herein, each standalone DMD frequency represents a group of fluid particles travelling at similar convective speeds or eddies forming in similar sizes. Therefore, the DMD decomposition echoes with the Richardson-Kolmogorov decomposition of turbulence into eddies of difference wavenumber and the process of energy cascade [30].

## 3. Numerical Details and Validation

We present the numerical details and validations of our LES in this section. The purpose is to assure the subsequent DMD analysis is driven by accurate data, eliminating the possibility of erroneous input. To main concision, we direct readers to Appendix I for the mathematical formulation of the Large-Eddy Simulation with Near-Wall Resolution.

### 3.1 Numerical Domain

The ensuing **Fig.** 2 illustrates the computational domain and the boundary conditions prescribed to our LES-NWR. The configuration is a classic turbulent, free-shear flow around an infinite square prism with side length *D*. The inflow, with the free-stream velocity $U_\infty$, has a uniform profile and is free of initial perturbation. We replicated the dimensions of the inlet, outlet, and laterals from the DNS domain in Portela *et al.* [41]. Yet, we employed a spanwise dimension of *4D* for numerical ease in filtering in the physical (Cartesian) space, whereas [41] used *πD* for that in the wavenumber space. For this particular flow during the shear layer transition (*Re=22,000*), *4D* suffices and has been adopted by most works in the literature [29].

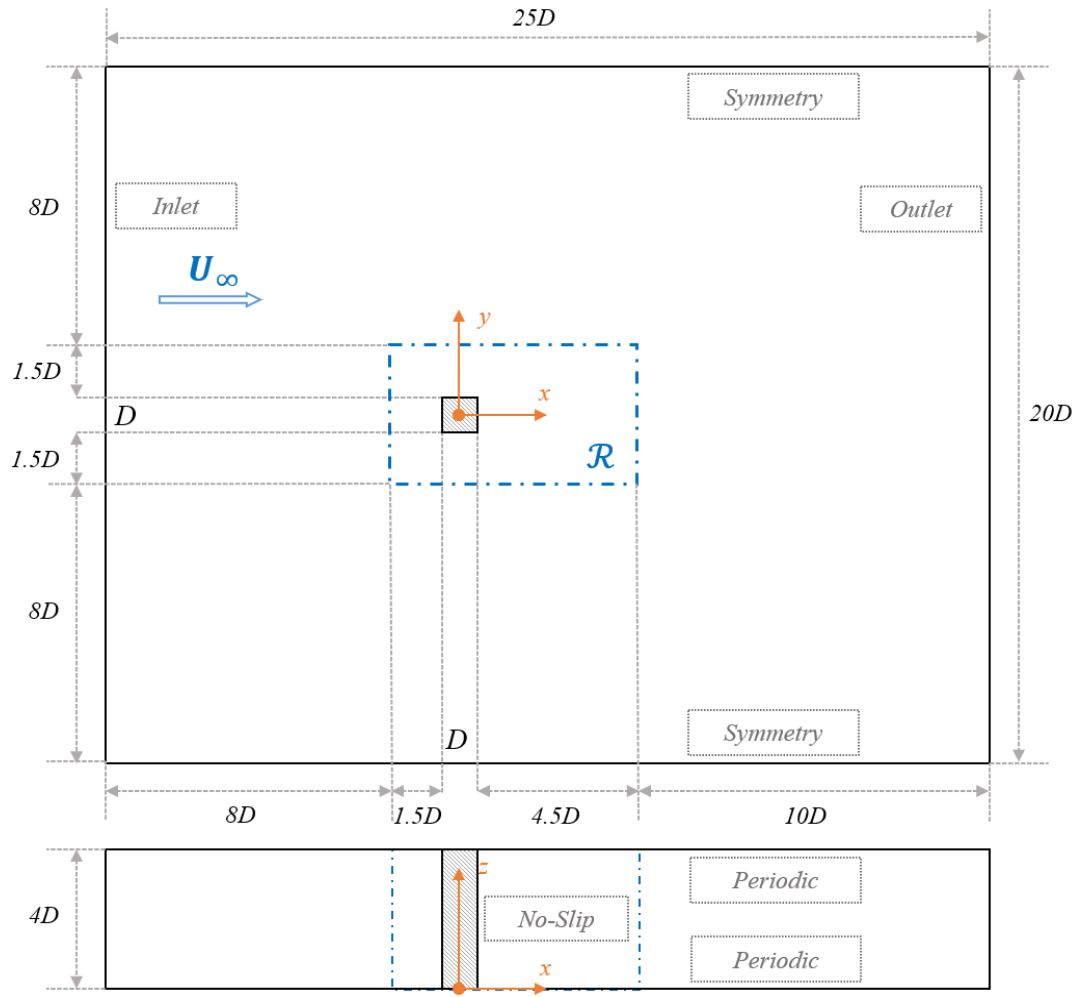

**Fig. 2** A schematic illustration of the computational domain and boundary conditions.

However, readers are reminded that during the 2D-to-3D transition of the prism wake between *Re=150-250*, modes A, B, and S produce spanwise vortices as large as *5.2D*, *1.2D*, and *2.8D*, respectively. Interested readers may refer to [29], [42]– [44] for details. Likewise, in the high-*Re* subcritical regime, another type of spanwise vortex occurs in scales as large as *9D* and *14D* in circular cylinder and prism wakes, respectively [45], [46]. This vortex is unique to ultra-slender structures and therefore beyond the scope of the present work. Nevertheless, we reiterate the importance of an appropriate spanwise dimension for different configurations of interest.

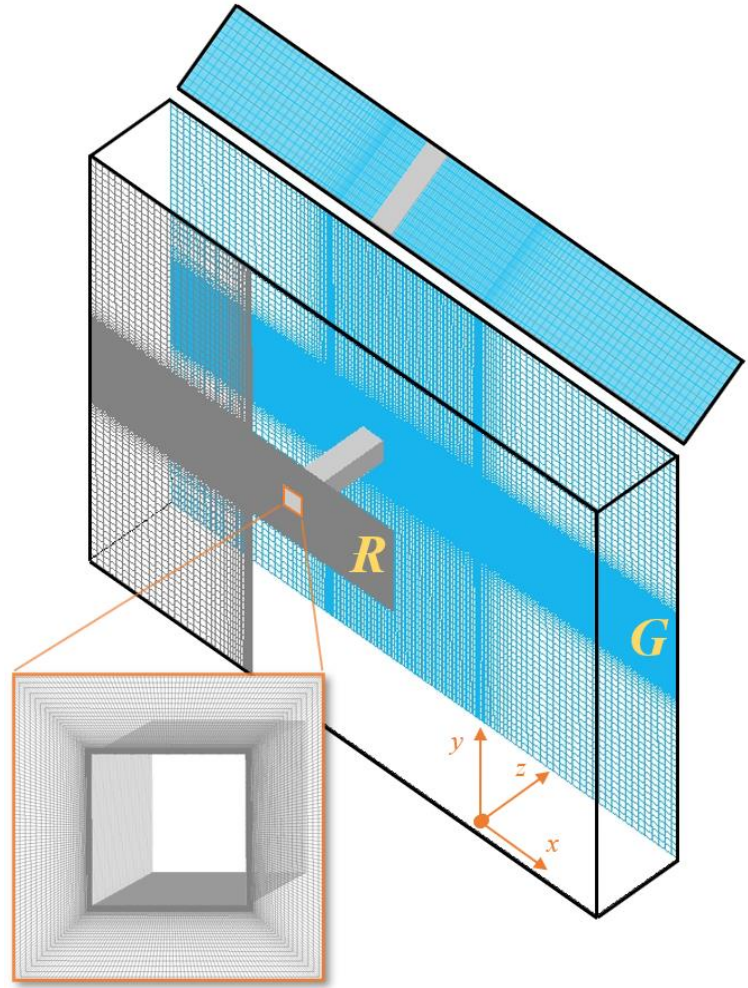

**Fig. 3** A hexahedral and non-conformal grid of 4.3 million elements.

Our LES-NWR employed a strictly hexahedral grid to minimize numerical errors (**Fig.** 3). The grid *G* consists of *4.3* million elements in sum and *2.9* million elements in a refinement region *R* around the prism. *R* always encloses the shear layers and near wake, thus capturing all the global phenomena and local extremities of turbulence. Four regions with coarser elements, namely the inlet, the outlet, and the two laterals adhere to *R* in a non-conformal fashion, as described by [47]. The non-conformal griding significantly lowered the mesh size while preserving high accuracy (shown later). **Table** 2 summarizes the details of the grid *G*.

**Table 2** Details of the hexahedral, non-conformal grid $G$.

| $\Delta x_{min}$ | $\Delta y_{min}$ | $\Delta z$ | $N_R$ | $N_G$ |
|---|---|---|---|---|
| $^3/_{400}\,D$ | $^1/_{4000}\,D$ | $^1/_{20}\,D$ | 2,903,400 | 4,287,864 |

We estimated the thickness of the viscous sublayer to determine the height of the wall-adjacent cells. The estimation relied on an analogy of the laminar boundary layer over a flat plate, as pointed out by White [48]:

$$\delta_{BL,99\%} \approx 3.5 \sqrt{\frac{2\nu\,l_x}{U_\infty}} \approx \frac{5.0\,l_x}{\sqrt{Re_x}} = 0.024D \tag{3.1.1}$$

where $l_x$ denotes the characteristic length of the boundary layer, which is taken as $0.5D$ after Cao *et al.* [49]. We set a fine resolution of $^1/_{4000}\,D$ to keep the $y^+$ strictly under unity. We also set a grading ratio of *1.05* to resolve the viscous sublayer by as many as *36* layers. Finally, we followed Menter [50] for suggested resolution in the *x*- and *z*-directions.

## 3.2 Numerical Methods

We employed a finite-volume, segregated, pressure-based solution algorithm for this low-Mach-number incompressible flow. The projection-based method obtains the velocity field from the momentum equation and satisfies the continuity by corrections of the pressure equation. To this end, we chose the Pressure-Implicit with the Splitting of Operators (PISO) scheme for the pressure-velocity coupling. The neighbor and skewness corrections of the PISO, particularly for transient simulations, notably improve the efficiency and robustness of the original SIMPLE-based solver. We also chose second-order schemes for the pressure interpolation and the spatial discretization of the viscous term, and the second-order bounded central-differencing scheme for that of the convection term.

In temporal discretization, we selected the bounded second-order scheme for time integration. We set a small time-interval $\Delta t^*$, such that the Courant-Friedrichs-Lewy (CFL) convergence condition is always satisfied, which eliminates the time marching issues in solving partial differential equations:

$$\Delta t^* = \frac{\Delta t \, U_\infty}{D} = 1.61 \times 10^{-3}$$

$$(3.2.1)$$

We sampled data at $t^* = 5\Delta t^*$ and only in $R$ to expedite the simulation and avoid excessive data storage. Finally, we selected the least-squares method for the evaluation of gradients and derivatives in the post-analysis.

## *3.3 Statistical Stationarity*

In turbulent flows, the so-called steady state is not an entirely appropriate description. The ever-changing nature of turbulence makes two identical flow snapshots nearly impossible, let alone the predictability implied by the term *steadiness*. In practice, one may only identify a 'steady state' in the statistical descriptions of turbulence called the statistical stationarity.

### 3.3.1 Global Statistics

First, we eyed on the global fluctuating lift coefficient for stationarity (**Fig.** 4). Despite persistent fluctuations due to unsteady motions, the root-square-mean (RMS) and the mean lift reached statistical stationarity before $1.2 \times 10^5 t^*$. Consequently, we began sampling for the DMD analysis thereafter. We sampled a range of *24* oscillation cycles at the highest frequency $1/t^*$, which pushed the storage limit of our HPC server by amassing more than $2.3 \times 10^4$ field snapshots.

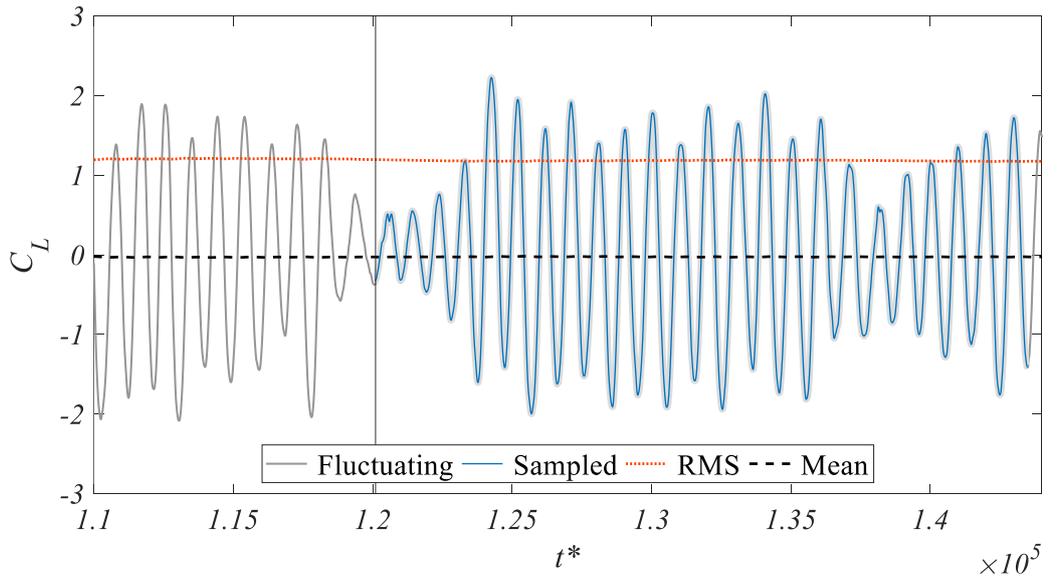

**Fig. 4** Time histories of instantaneous, RMS, and mean lift coefficients. The DMD sampling range consists of 24 oscillation cycles in the statistical stationary state.

### 3.3.2 Local Statistics

We further assured the statistical stationarity by local statistics. **Fig.** 5a illustrates the seven nodes selected to monitor the mean local velocities. With respect to the prism, the nodes are characteristic of the inflow, the pulsating shear layers, and the turbulent wake. At all monitor points, vast instabilities are observed in the early and transitional stages. By contrast, the normalized mean velocities exhibited stationarity by approaching clear asymptotes before $1.0 \times 10^5\ t^*$, reaffirming the observations on global statistics.

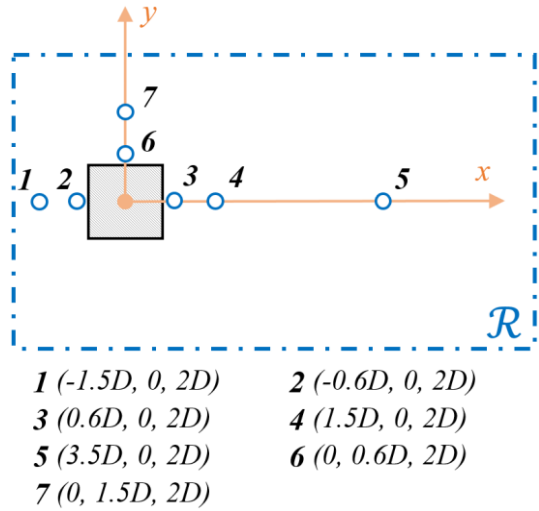

*1* *(-1.5D, 0, 2D)*     *2* *(-0.6D, 0, 2D)*
*3* *(0.6D, 0, 2D)*      *4* *(1.5D, 0, 2D)*
*5* *(3.5D, 0, 2D)*      *6* *(0, 0.6D, 2D)*
*7* *(0, 1.5D, 2D)*

a)

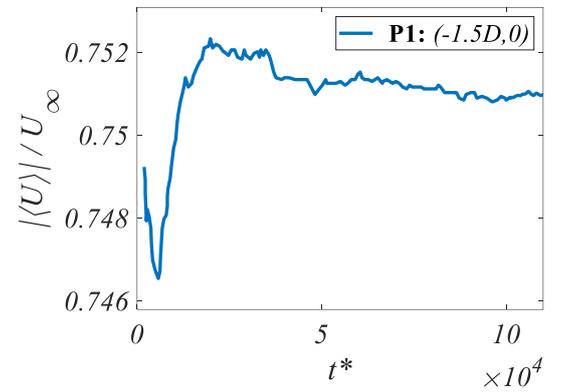

b)

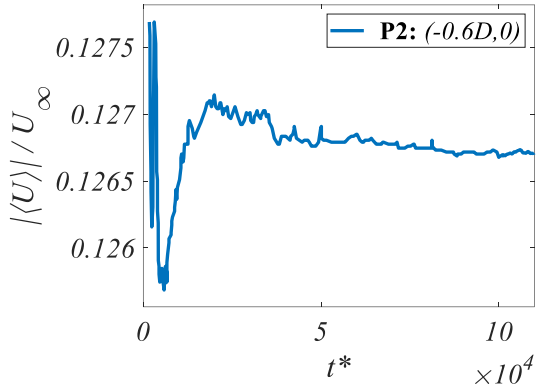

c)

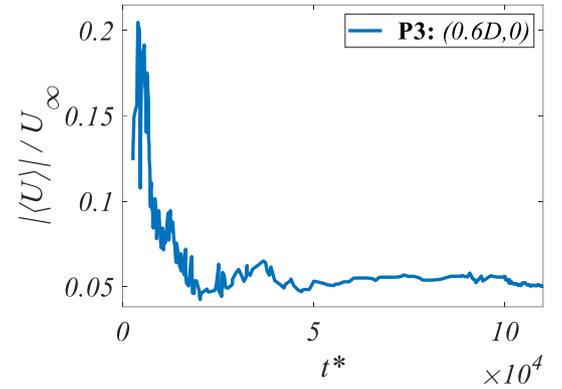

d)

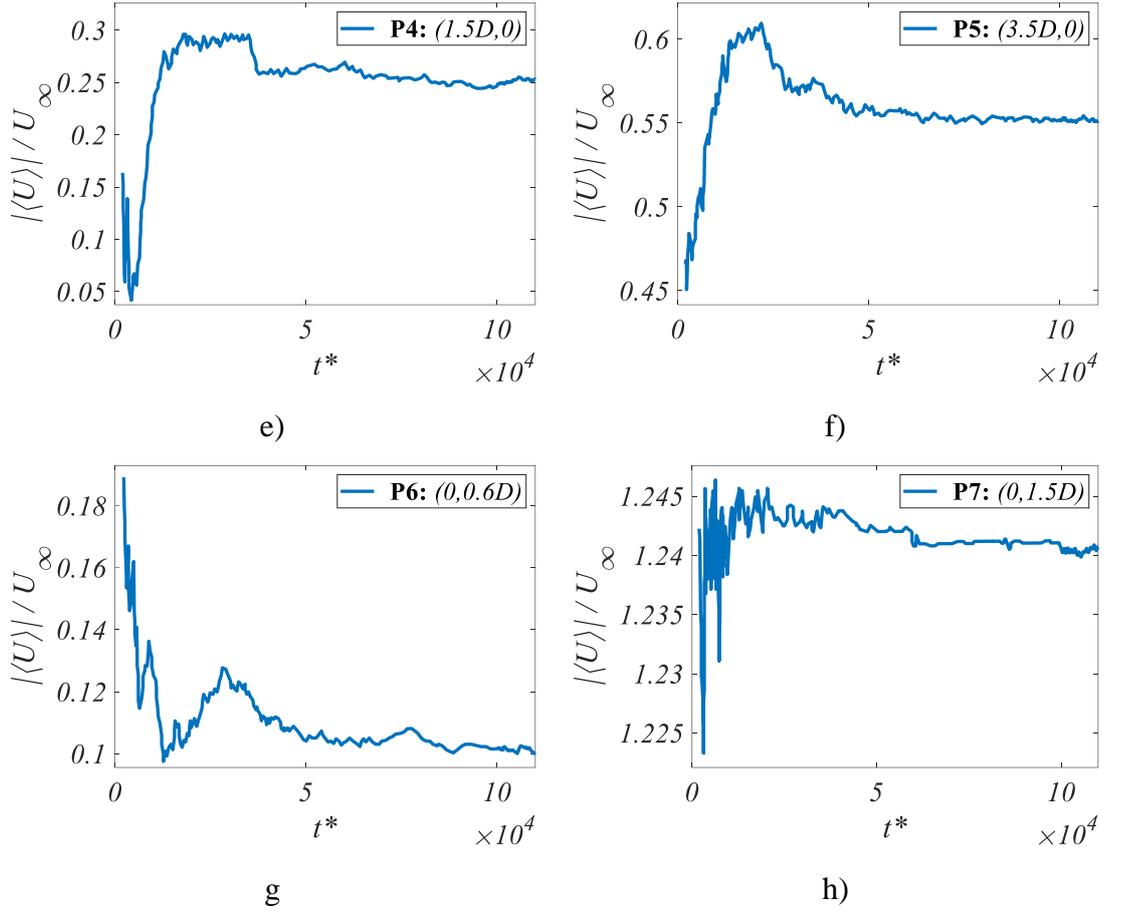

**Fig. 5** a) A schematic illustration showing the locations of monitor points 1-7; b)-h) normalized mean velocity magnitude versus time step at monitor points 1-7, showing statistical stationarity.

### *3.4 Grid Resolution*

The resolution of our structured grid $G$ is critically important to the accuracy of LES-NWR, so we adopted stringent criteria for its assessment.

### 3.4.1 Resolved Spectrum

Recall the definition of LES-NWR, a proper implementation shall resolve all the large-scale eddies in the energy-containing range and only model the smaller, more trivial ones towards the dissipation range. This is to say that, if an LES filter sits inside the inertial subrange, then the Navier-Stokes equations guarantee the accuracy of large-scale turbulence. Therefore, grid resolution is often the most indicative evidence of LES accuracy. In his masterpiece on turbulence, Pope [30] defined a demarcation between the energy-containing range and the inertial subrange -- 80% resolution for the total turbulence kinetic energy (TKE).

To quantify the resolved portion, we denote the resolved TKE by $k_r$, the subgrid TKE by $k_{sgs}$, and the numerical TKE by $k_{num}$. We then express the resolved spectrum, $E$, by:

$$E \equiv G(x) \, E_{all} \approx \frac{k_r}{k_{all}}$$

*(3.4.1.1)*

where $k_{all}$ denotes the total TKE:

$$k_{all} = k_r + k_{sgs} + k_{num}$$

*(3.4.1.2)*

$$k_r = \frac{1}{2}\left(\overline{u'^2} + \overline{v'^2} + \overline{w'^2}\right)$$

*(3.4.1.3)*

$$k_{sgs} = v_{sgs}^2 / l_s^2$$

*(3.4.1.4)*

where $G(x)$ denotes the filter function in three-dimensional space; $\overline{u'^2}, \overline{v'^2}$, and $\overline{w'^2}$ denote variance of the fluctuating velocities $u'$, $v'$, and $w'$, respectively. $k_{num}$ is a pseudo-energy term that accounts for discretization error and numerical residual. Celik *et al.* [51] pointed out that $k_{num}$ is sufficiently small for a LES-NWR with an overall second-order discretization. As introduced before, our discretization is at least second-order, producing minimal numerical dissipation. As we also strictly maintained the Courant-Friedrichs-Lewy (CFL) condition, the numerical dispersion is insignificant. Our convergence criteria are also stringent, at $1 \times 10^{-6}$, for both the continuity and momentum equations. Therefore, we treated $k_{num}$ as negligible.

The ensuing **Fig. 6** presents the resolved spectra of the mid-span x-y plane and the prism walls. Given the incompressible Newtonian fluid and a grid-dependent filter, $E$ is inversely proportional to the local velocities and cell volume. Evidently, our grid resolved at least 90% of the TKE. The same was true for the prism walls, except for some narrow strips near the corners A and B, where the resolved TKE is about 77%. We anticipated them because local accelerations occur as the result of sharp corner separation. Since TKE is proportional to local *Re*, the resolution requirement becomes more stringent near the corners. In general, the resolved spectra prove that our filter sits inside the inertial subrange, hence the resolution of our grid is suitable for LES simulation.

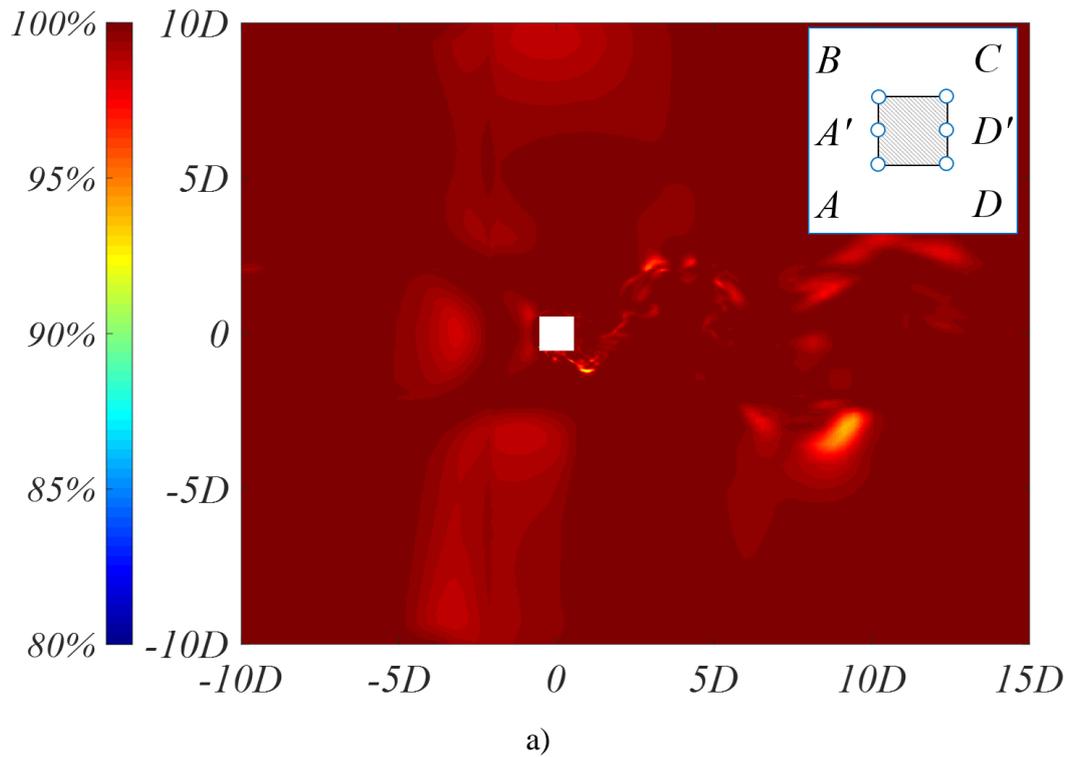

a)

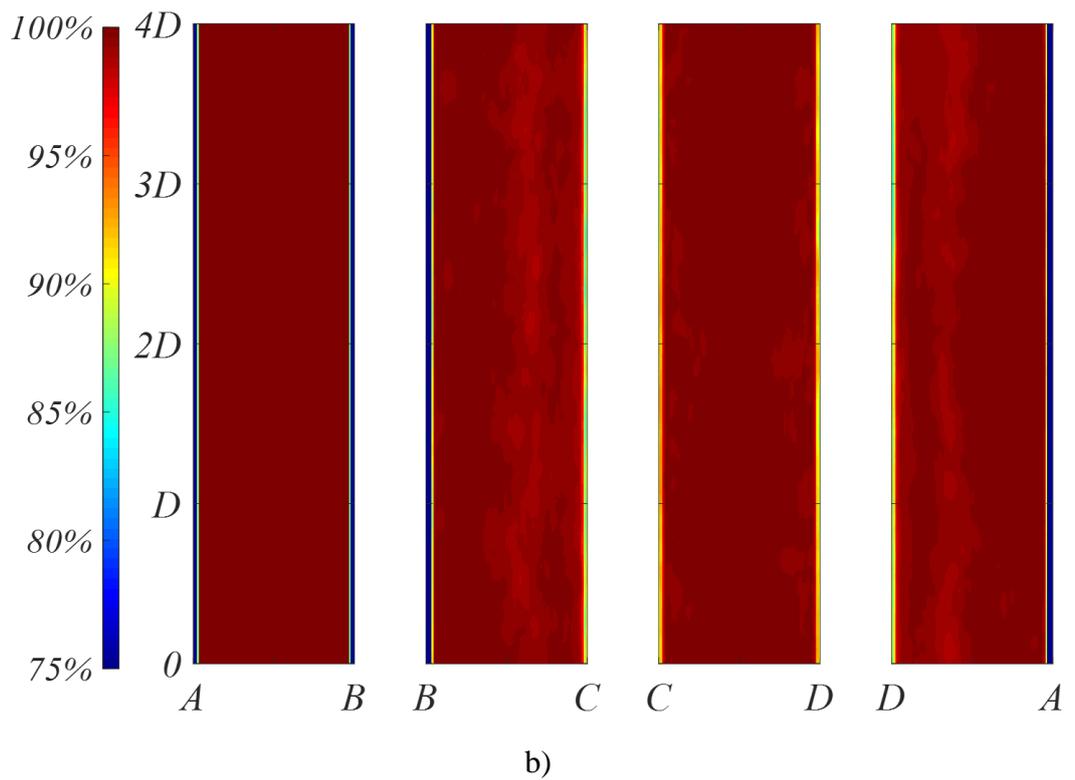

b)

**Fig. 6** Contours showing the time-averaged resolved spectra of the turbulence kinetic energy of a) the mid-span x-y plane and b) prism walls. An 80% resolves at least the energy-containing range.

### 3.4.2 LES_IQ Index

In the fluid mechanics community, some researchers, like Davidson [52], had challenged the 80% demarcation. Though this topic is beyond our scope, we still looked for an alternative assurance to our grid resolution. Celik *et al.* [51] proposed the *LES_IQ*, an index quantifying the resolution of a LES grid, as:

$$LES_{IQ} = \left[ 1 + a_v \left( \frac{\nu_{sgs} + \nu}{\nu} \right) \right]^{-a_e}$$

(3.4.2.1)

where $\alpha_v = 0.05$ and $a_e = 0.53$ are empirically derived constants. Accordingly, *LES_IQ* < 80% indicates a Very Large-Eddy Simulation (VLES), *LES_IQ* > 80% signals a proper LES-NWR, and *LES_IQ* > 95% suggests a Direct Numerical Simulation (DNS).

Accordingly, **Fig.** 7 presents the *LES_IQ* contours of the mid-span x-y plane and the prism walls. Our grid achieved the DNS resolution in most fluid domains and on all prism walls. It also achieved at least the LES resolution in all fluid domains of interest, including the shear layers and the near wake. Although a few patches of VLES resolution existed in the far wake, but being neither influential to the upstream activities nor partial to the sampled domain, they are of trivial importance. Overall, the *LES_IQ* confirms the quality of our grid resolution.

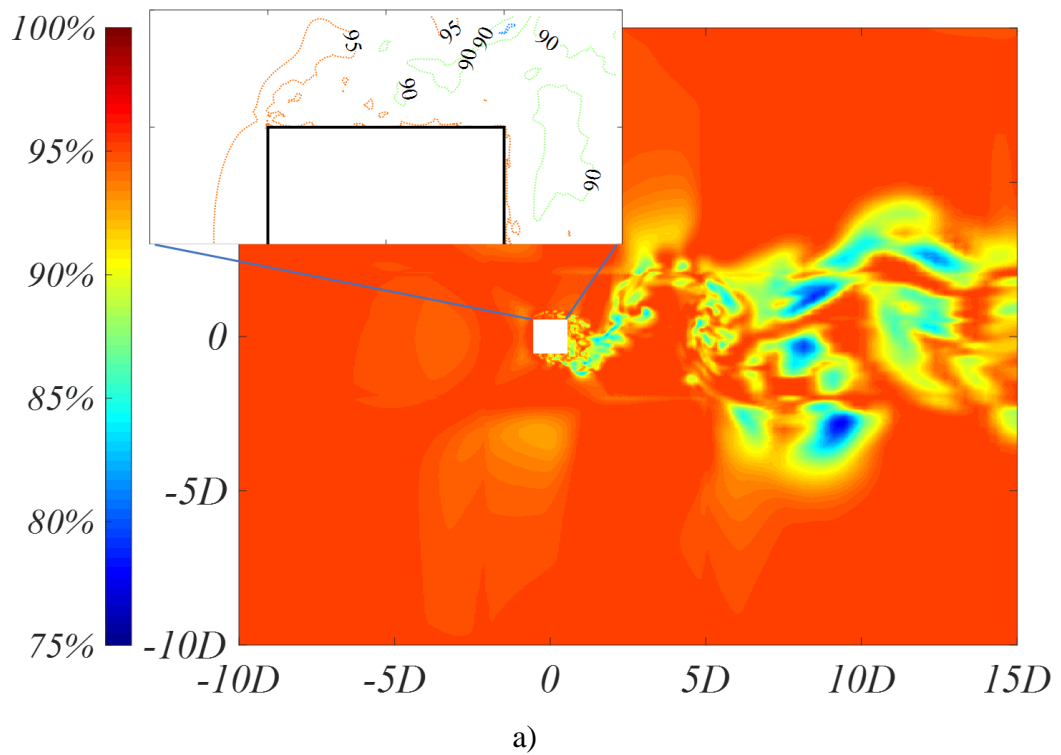

a)

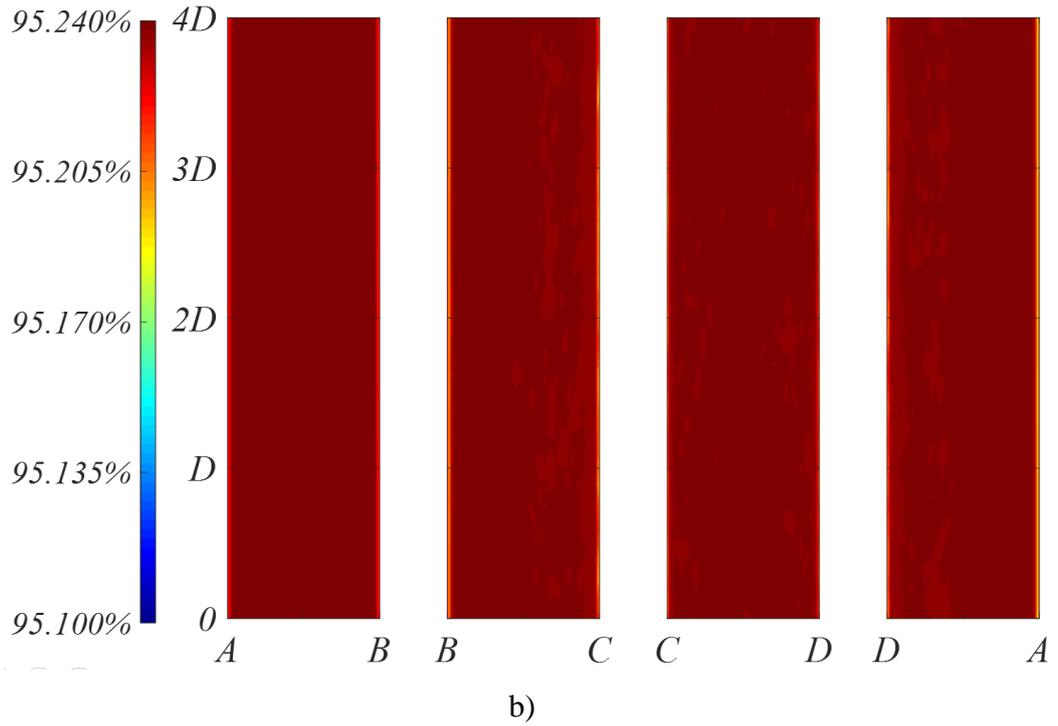

b)

**Fig. 7** Contours showing the *LES_{IQ}* index proposed by Celik *et al.* [51] of a) the mid-span x-y plane and b) prism walls. 80% signals satisfactory LES resolution.

## *3.5 Simulation Accuracy*

Since numerical accuracy is indispensable to our subsequent DMD analysis, beside the grid resolution, we compare the simulated results to the literature for further validation.

### 3.5.1 Global Statistics and Spectral Density

Below, **Table** 3 summarizes global force coefficients and the Strouhal number *St* of our simulation compared to previous experiments. The statistics not only agree with the experiments performed at exactly *Re=22,000,* but also an array of others in the subcritical regime, because one expects the universal occurrence of the shear layer transition II and an asymptotic convergence vortex formation length in this flow regime [29].

**Table 3** Global force coefficients and the Strouhal number compared to the literature

| $Re\ (10^3)$ | $\langle C_D \rangle$ | $C_{D,\ RMS}$ | $C_{L,\ RMS}$ | $St$ | Contribution |
|---|---|---|---|---|---|
| **22** | **2.048** | **0.200** | **1.173** | **0.127** | **Present work** |
| 22 | 2.069 | 0.146 | 1.221 | 0.126 | Li *et al*. [8] |
| 100 | 2.05 | 0.17 | 1.3 | 0.12 | Vickery [53] |
| 176 | 2.04 | 0.22 | 1.19 | 0.122 | Lee [54] |
| 22 | 2.1 | - | 1.2 | 0.13 | Bearman/Obasaju [55] |
| 27 | 1.9-2.1 | 0.1-0.2 | 0.1-0.6 | - | Cheng *et al*. [56] |
| 23 | 1.9-2.1 | 0.1-0.2 | 0.7-1.4 | - | McLean/Gartshore [57] |
| 22 | 2.10 | - | - | 0.130 | Norberg [58] |
| 34 | 2.21 | 0.18 | 1.21 | 0.13 | Luo *et al*. [59] |
| 21.4 | 2.1 | - | - | 0.132 | Lyn *et al*. [60], [61] |

On the other hand, we determined the Strouhal number *St* by performing the Fourier transformation and power spectral density analysis for the fluctuating global lift coefficient, $C_L^{'}$. Apart from the stellar agreement in global *St*, the periodogram (**Fig. 8**) supports our previous conclusions on the grid resolution. The curve lucidly exhibits the Kolmogorov *-5/3* Law before *St=1.1*. As a fundamental pillar to the Richardson-Kolmogorov energy cascade, the -5/3 power-law is unique to the inertial subrange and often regarded as its most indicative icon [30]. Therefore, its appearance signifies that at least a significant portion of the inertial subrange was resolved by the Navier-Stokes equations based on our grid *G*. On a different note, one shall expect an exponential decay in the dissipation range of the full turbulence spectrum, instead of the linear decay herein towards the high-frequency space. Nevertheless, we find the observed linearity truthful to the subgrid dynamics, especially considering the one-equation, linear mixing-length hypothesis undertaken by the Smagorinsky model. The linearity shows the filtering process only takes place inside and beyond the inertial subrange, depicting the anticipated spectrum of a proper LES-NWR.

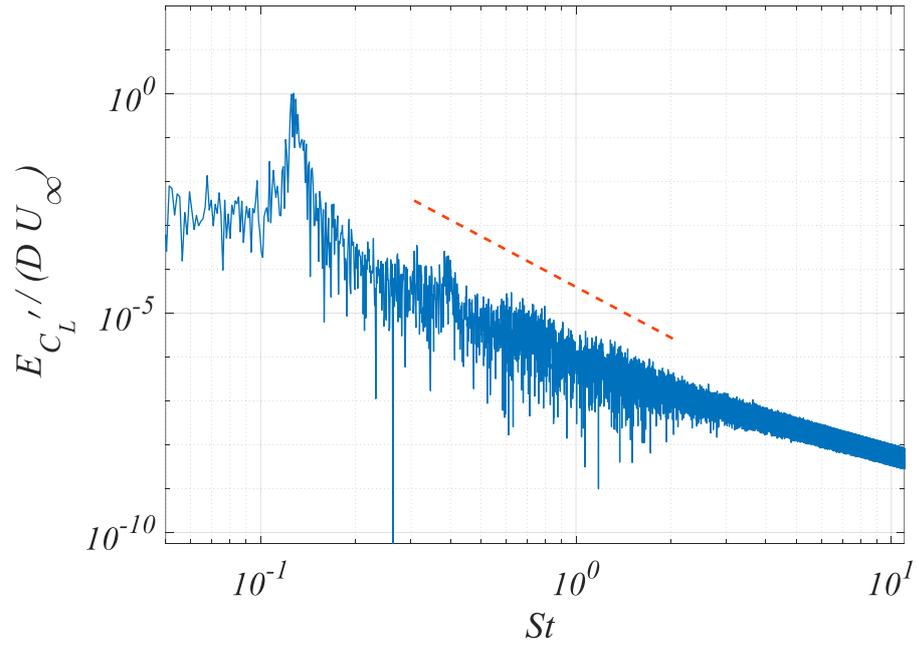

**Fig. 8** Periodogram of the normalized spectrum of the fluctuating global lift coefficient, showing the Strouhal number for the Kármán vortex shedding and an agreement with the Kolmogorov - 5/3 Law [30].

### 3.5.2 Prism Walls

Next, we examine the grid resolution at the prism walls for validation. **Fig.** 9 presents the $x^+$ and $y^+$, or the dimensionless wall distances, calculated from wall shear and the notion of friction velocity. Our grid met the $x^+\approx30$ and $y^+\approx1$ requirements proposed by Menter [50]. The $z^+$ is trivial in this infinite length configuration. Moreover, we compared the time-averaged, normalized pressure coefficient on the prism walls to the literature (**Fig.** 10). Our results agree well with a range of wind tunnel and DNS studies, especially the more recent empirical results like Nishimura [62] and Nishimura and Taniike [63]. Notably, some variations occur on the downstream wall across the existing studies. Nonetheless, the present result falls well within the comparative range of wind tunnel and DNS studies, validating the accuracy and fidelity of our LES-NWR.

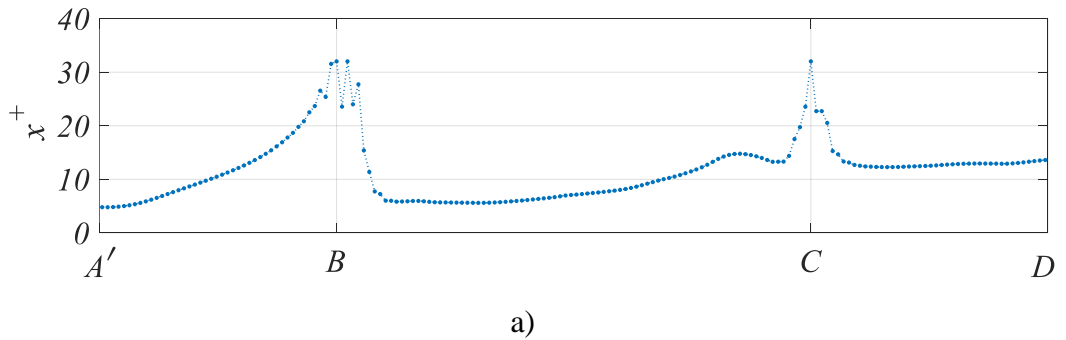

a)

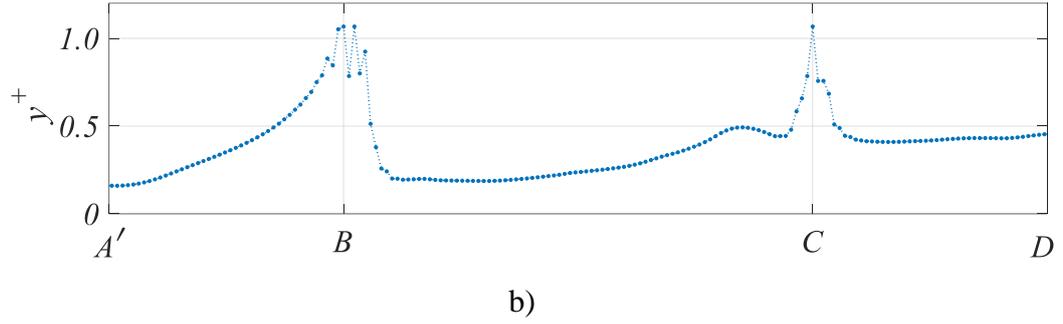

b)

**Fig. 9** Time-averaged values of a) $x^+$ and b) $y^+$.

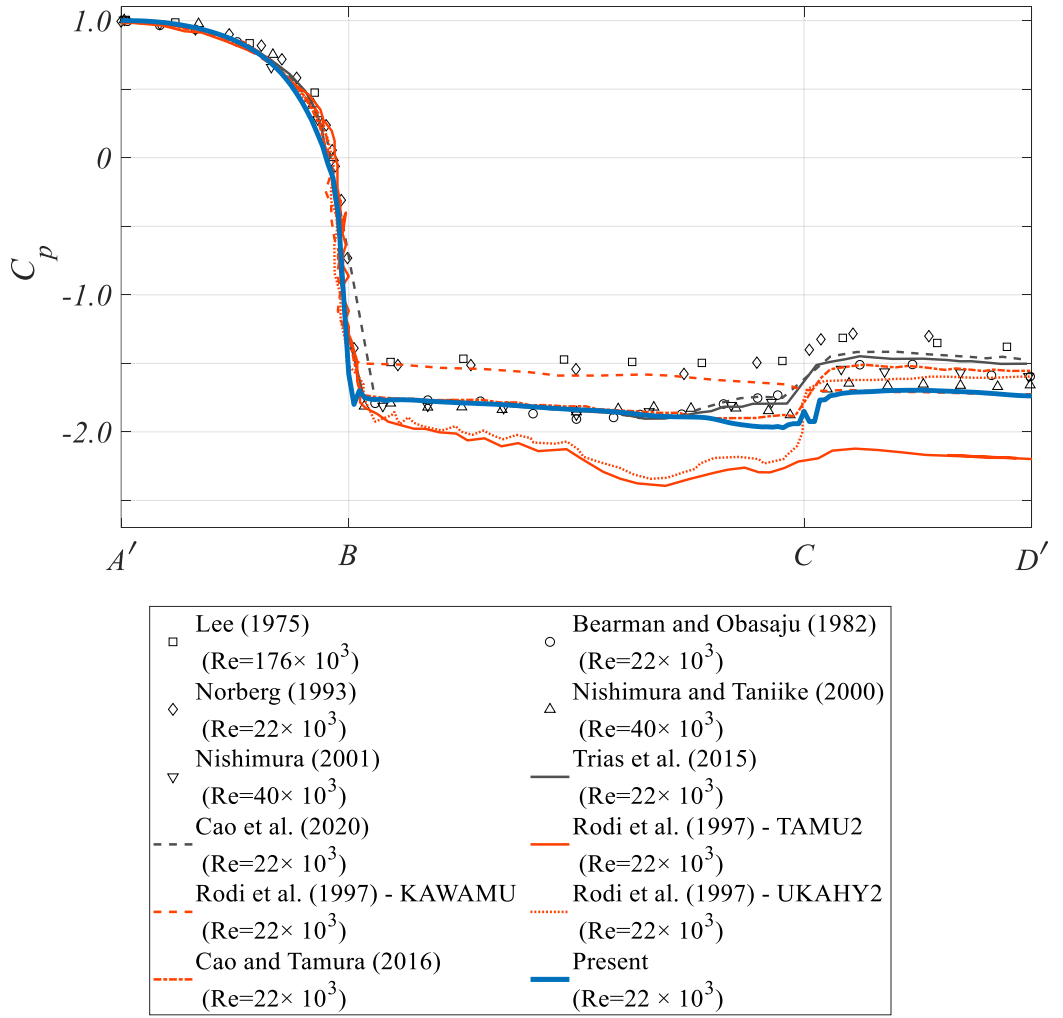

**Fig. 10** Time-averaged and normalized pressure coefficient of prism walls compared to the literature. Markers: experiments. Curves: black-DNS, orange-LES, blue-present work.

### 3.5.3 Velocity Field

As the final piece of validation, we examine the velocity field. **Fig.** 11 presents the time-averaged, normalized $u$ along the zero-ordinate. Our results sit fittingly among the array of experiments and DNS studies. To ensure such an agreement is not fortuitous, we also examine the fluctuating velocities that characterize turbulence. **Fig.** 12 presents the time-averaged, normalized *u', v',* and *w'* along the zero-ordinate. Besides the self-evident validation of our results, we also made an interesting observation between the experiments and simulations.

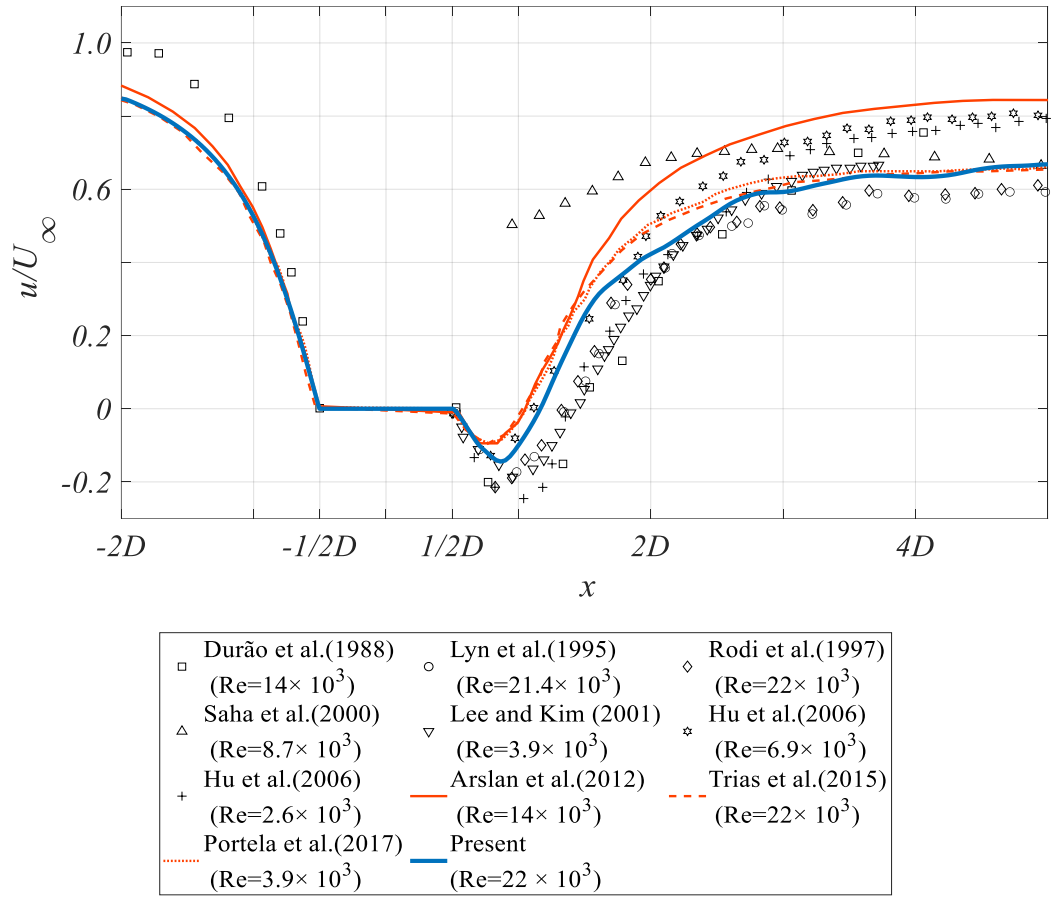

| | Durão et al.(1988) (Re=14× $10^3$) | ∘ | Lyn et al.(1995) (Re=21.4× $10^3$) | ◇ | Rodi et al.(1997) (Re=22× $10^3$) |
|---|---|---|---|---|---|
| △ | Saha et al.(2000) (Re=8.7× $10^3$) | ▽ | Lee and Kim (2001) (Re=3.9× $10^3$) | | Hu et al.(2006) (Re=6.9× $10^3$) |
| + | Hu et al.(2006) (Re=2.6× $10^3$) | | Arslan et al.(2012) (Re=14× $10^3$) | | Trias et al.(2015) (Re=22× $10^3$) |
| | Portela et al.(2017) (Re=3.9× $10^3$) | | Present (Re=22 × $10^3$) | | |

**Fig. 11** Time-averaged and normalized *u* along the zero-ordinate compared to the literature. Markers: experiments. Curves: orange-DNS, blue-present work

For fluctuating velocities, unlike the instantaneous velocity ***u***, high-fidelity simulations (DNS and proper LES-NWR) tend to agree closely with each other and only moderately with the experiments. The experimental results are somewhat scattered even among themselves. Although one can hardly judge the right and wrong, we reflect on the necessary symbiosis between experiments and numerical simulations. Since the former is the most direct portrayal of physics, one often sees it as the unwavering shrine of truth. However, for many elusive phenomena like turbulence, limitations in apparatus or simply in the number of sampling points may cause perspectives partial to the whole. With the potentials of quantum computing, so the foreseeable ubiquity of LES-NWR or even DNS in the near future, we see the CFD as the finest wingman, if not the missing puzzles, for empirical investigations in addition to experiments. The duo will elevate physical insights and engineering applications to unprecedented heights.

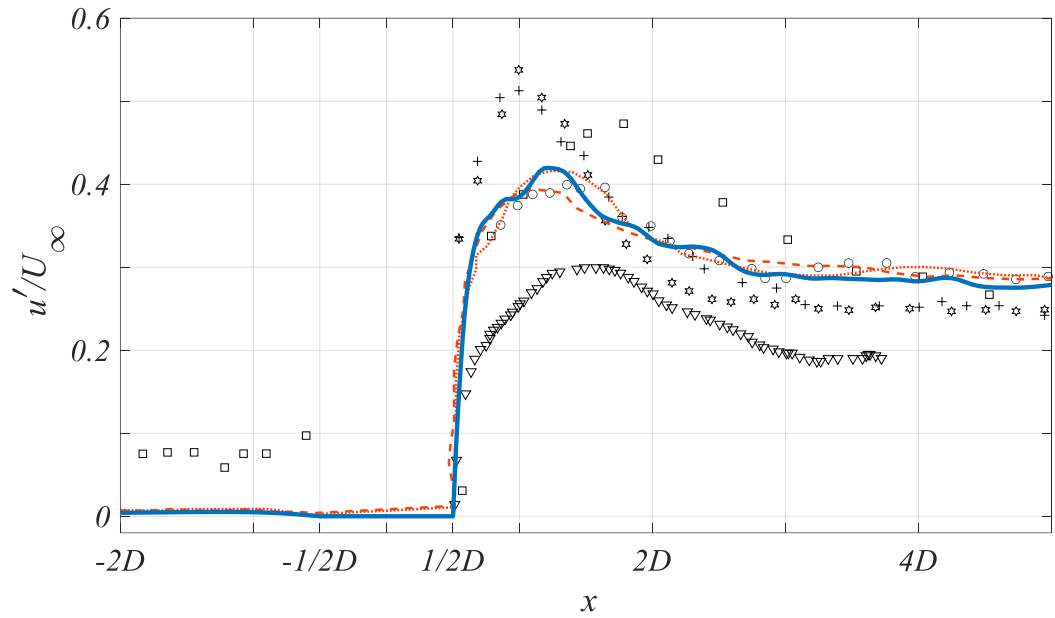

a)

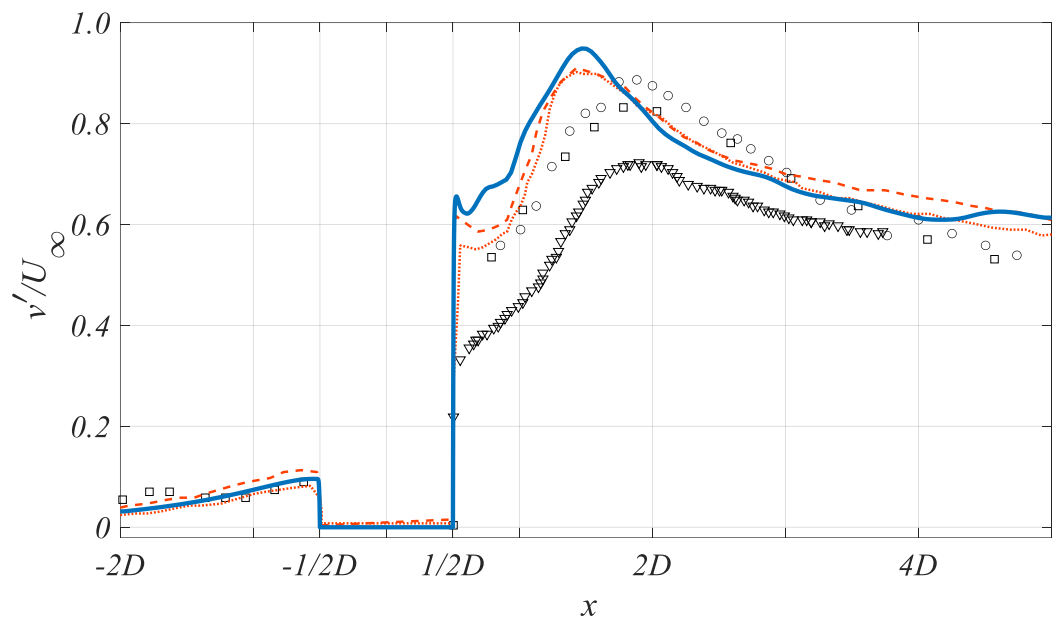

b)

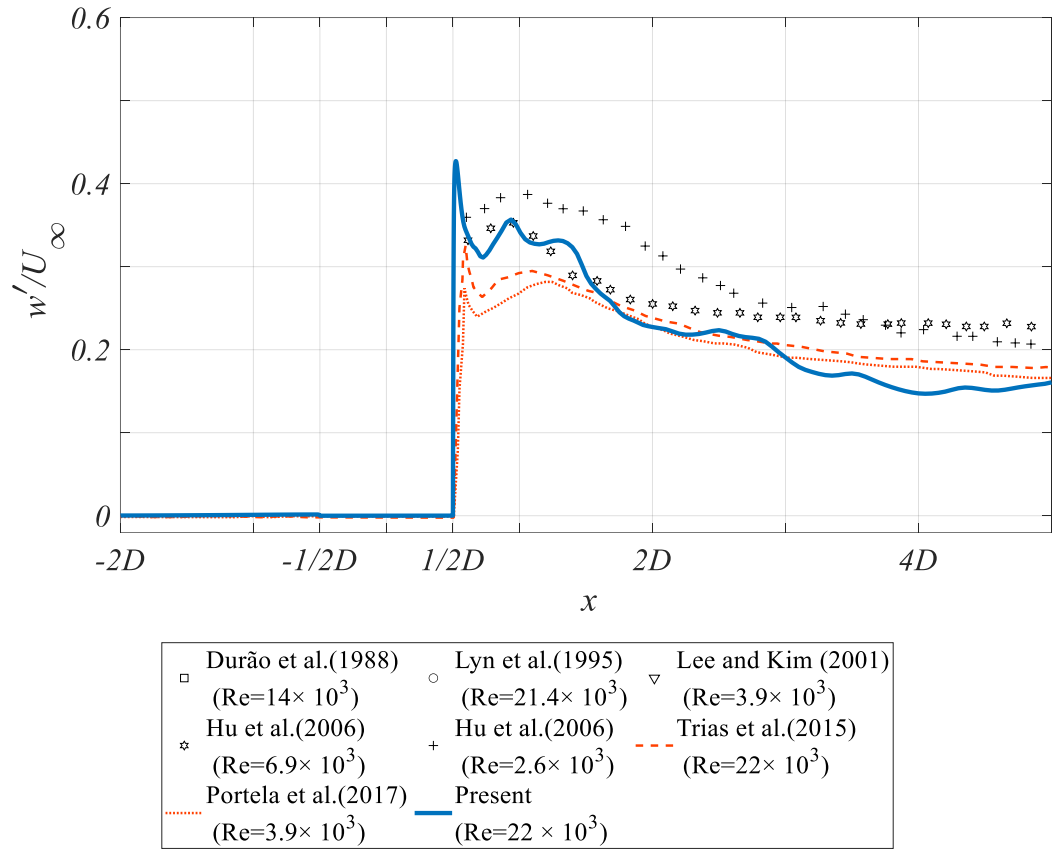

c)

**Fig. 12** Time-averaged and normalized fluctuating a) $u$, b) $v$, and c) $w$ along the zero-ordinate compared to the literature.

# 4. Benchmark on Sampling Range

The sampling range and resolution are the two independent variables in this parametric investigation of DMD sampling. Before assessing their combined effect, we establish a benchmark by parametrizing only the sampling range while keeping the sampling resolution constant and at its highest frequency. The convergence of the sampling range indicates that of the Krylov sequence, which signals minimal ensemble differences, or time-mean differences in the statistical stationary case, with data repetition [16].

## 4.1 Convergence and Modal Frequency

We quantify the sampling range by the number of cycles--an intuitive measure for this statistically stationary, oscillatory, and turbulent free shear flow. **Fig.** 13 presents the Strouhal numbers *St* of the most dominant DMD modes 1-3 versus the number of oscillation cycles. Clearly, the modes exhibit different degrees of fluctuation below cycle 10. By contrast, they universally stabilize above cycle 11, tending asymptotically towards the *St = 0.127, 0.121*, and *0.006*, for the DMD modes 1, 2, and 3, respectively. On this note, for some modes like mode 1, fluctuations are sufficiently small even after cycle 5. We point out that the mechanism portrayed by mode 1 may be vastly different from those by the others, so we try to make the best judgment on the global state $\tilde{A}$ instead of over-relying on the behavior of a single mode. Overall, **Fig.** 13 lucidly demonstrates the convergence of the DMD with increasing sampling range up to 20 cycles, and the sudden divergence in the subsequence.

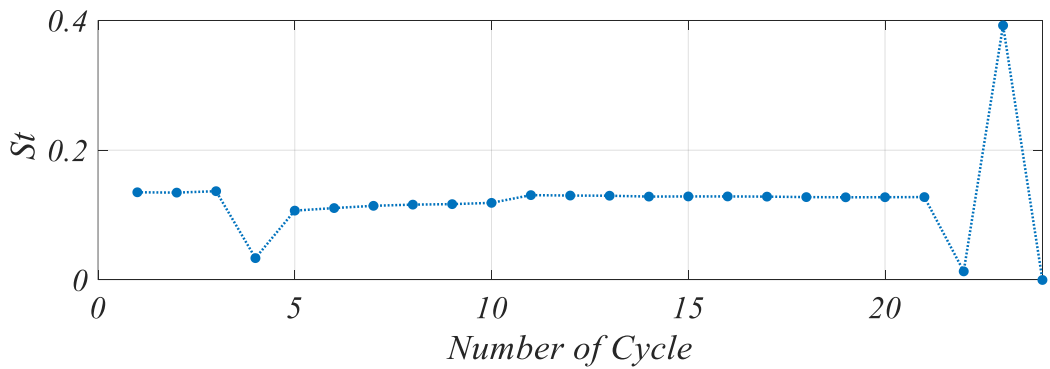

a)

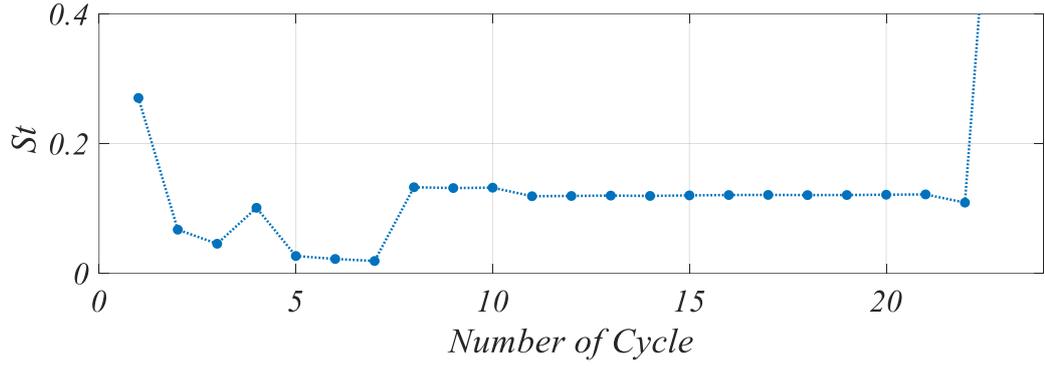

b)

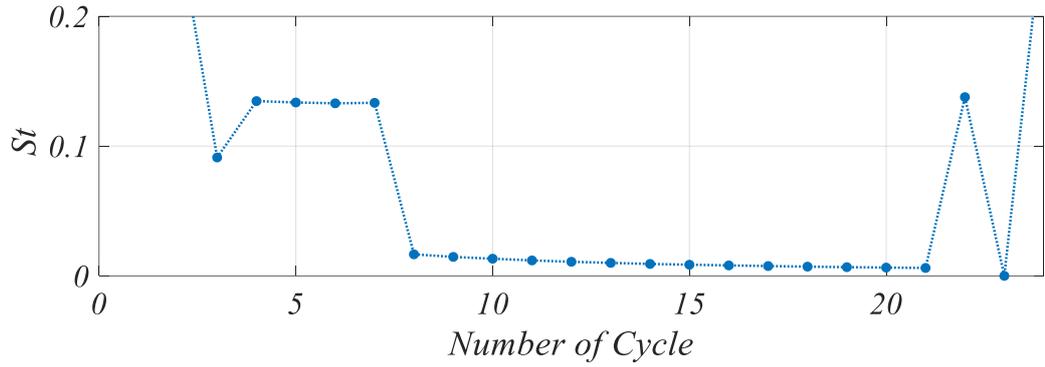

c)

**Fig. 13** The Strouhal number *St* versus the number of DMD-sampled oscillation cycles of dominant DMD a) mode 1, b) mode 2, and c) mode 3.

### 4.2 Convergence and Stability

On **Fig.** 13, we take note of an unanticipated behavior towards the high-cycle end —the convergence deteriorates drastically and almost universally on cycle 22 and beyond. The observation contradicts the logical intuition that convergence becomes progressively better, or at least consistent, with an increasing sampling range. This divergence also disobeys the predictions of previous investigations on sampling convergence [4], [12], [13]. To this end, we eye on the growth rate to analyze the perplexity. **Fig.** 14 displays a distinct constancy of the growth rate up to cycle 22. The infinitesimal magnitudes ($<\pm10^{-6}$) also appeal to excellent modal stability. On the zoom-in sub-figure, one may note that small fluctuations begin to build up between cycles 15 to 21, but the magnitudes are well within the $10^{-6}$ tolerance therefore negligible.

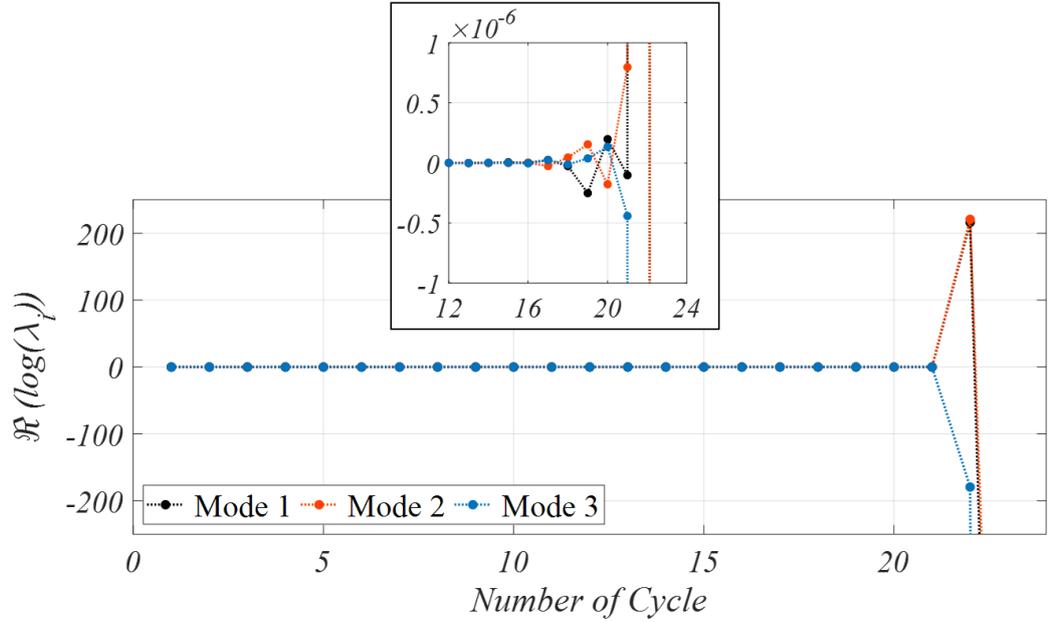

**Fig. 14** Growth rate versus the number of DMD-sampled oscillation cycles of dominant DMD modes 1-3.

What triggers the sudden, colossal (by nine orders of magnitude) divergence on cycle 22? We anticipate it is due to the loss of modal stability. To validate the notion, we examine the DMD spectrum. **Fig.** 15 presents the DMD spectra for cycles 1, 21, 22, and 24. In a perfectly oscillatory, or marginally stable system, the modes, or the poles of the system, represent singularities where the system behaves with regularity and will lie exactly on the $\Re^2 + \Im^2 = 1$ unit circle. As expected, the DMD modes for cycles 1 and 21, although not perfectly oscillatory, are infinitely close to the unit circle. The corresponding radius free of poles, or the Region of Convergence (ROC), also contains the $\Re^2 + \Im^2 = 1$, manifesting the system's stability. The ROC's inclusion of the origin also indicates an acausal, or anti-causal system, which means the system depends only on future input and not the past input. This is concrete evidence of a time-invariant Koopman description. On the contrary, the modal stability experiences a clear and drastic deterioration for cycle 22. The scattered poles lie far from and on either side of $\Re^2 + \Im^2 = 1$. The ROC also no longer encloses the unit circle, indicating the loss of system stability. A pole also appears near the origin, shrinking the ROC down to an annulus. Therefore, the system is no longer acausal. With more encroachment of the unit circle, the stability and causality further exacerbate for cycle 24.

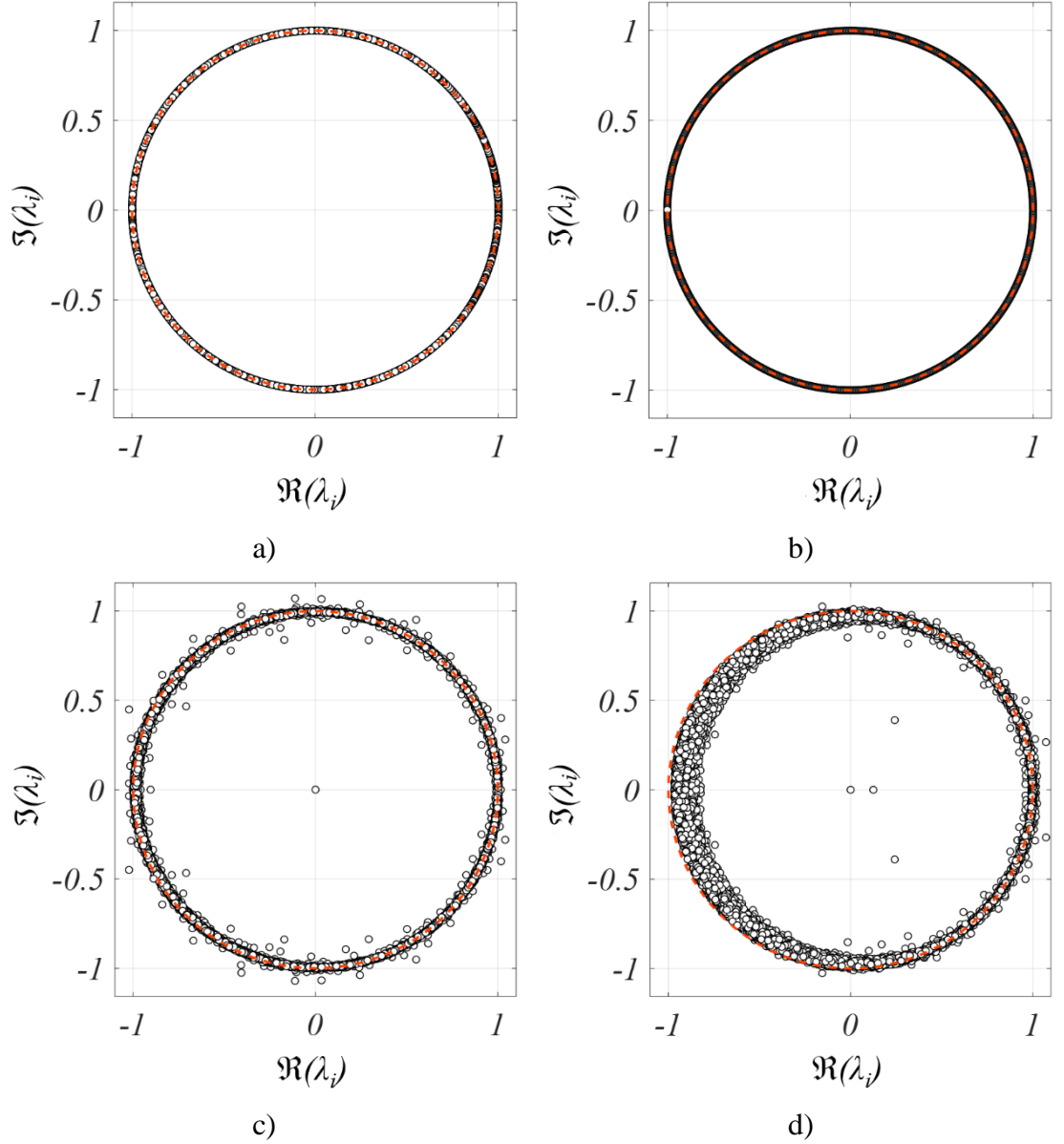

**Fig. 15** The DMD spectra of a) 1, b) 21, c) 22, and d) 24 DMD-sampled oscillation cycles.

Despite the observations, the origin of the instability might not be straightforward. With a well-established DMD algorithm, the likelihood of numerical fault is scant. From a fluid dynamics point-of-view, the turbulence characteristics of cycles 21 and 22 cannot be vastly different, especially given the flow's statistical stationarity. Thus, the only possibility lies in the nuances of sampling. After scrutiny, we found our grid $G$ yields a spatial dimension (*i.e.,* number of nodes)of $n=21,257$ for the tested plane. The 24-cycle sampling range yields a temporal dimension (*i.e.,* number of snapshots) of $m=23,400$. Here, readers are reminded of one tacit and often taken-for-granted assumption of the DMD, which is $m<<n$. The temporal dimensions of cycles 21 and 22 are $m_{21}=20,475$ and $m_{22}=21,450$, respectively. The stability overturn takes place precisely over the range in which $m$ transcends $n$, that is, between cycle 21 and 22. As such, we conclude the violation of $m<<n$ results in unstable and spurious DMD decompositions. Moreover, empirical observation of this parametric test reveals that although the $m<<n$ must be strictly met, the extent of $m<<n$ is not as substantial as we anticipated. For engineering practices, $m<n$ generally suffices for modal stability. The $m<n$ condition originates from the Singular-Value Decomposition (SVD). The violation of the condition produces zero singular values or principal components, null spaces, and nontrivial cokernels. If we consider the geometric implication of the SVD, it is the transformation of a Euclidian unit sphere into a hyper-ellipse. Violation of the condition is equivalent to forcing a stretch of the sphere into an ellipse along a zero principal semiaxis. As the result, the DMD becomes ill-conditioned, and its outputs are bound to diverge.

### 4.3 Convergence and Reconstruction

The analysis of the Strouhal number revealed the convergence with sampling range. The investigation of modal stability also disclosed the importance of meeting the $m<n$ condition. In this section, we augment existing findings with an examination on the reconstruction accuracy.

### 4.3.1 Reconstruction Accuracy

We quantify the reconstruction accuracy by the mean $l_2$-norm of reconstruction error:

$$\|e\|_{2,ins} = \frac{1}{n} \sum_{k=1}^{n} \left[ \left( \frac{x_{DMD,k,i} - x_{k,i}}{x_{k,i}} \right)^2 \right]^{1/2}$$

(4.3.1.1)

where $x_{k,i}$ is the original input data, and

$$x_{DMD,k,i} = \sum_{j=1}^{r} \phi_j \exp(\omega_j t_i^*) \alpha_j$$

(4.3.1.2)

is the DMD reconstructed data at node $k$ and instant $i$. **Fig.** 16 presents the reconstruction error versus time step $t^*$. Given the increasing temporal dimension $n$, we conveniently divided the cycles into four groups, Groups 1, 2, 3, and 4, consisting of 1-8, 9-14, 15-20, and 21-24 cycles, respectively. **Fig.** 16a illustrates the supreme reconstruction accuracy of Group 1, echoing with the findings in [8], [9]. The error is limited to *0.1* and has only a few singularities exceeding the threshold. Given the erratic nature of turbulence, the singularities are both expected and trivial to global performance.

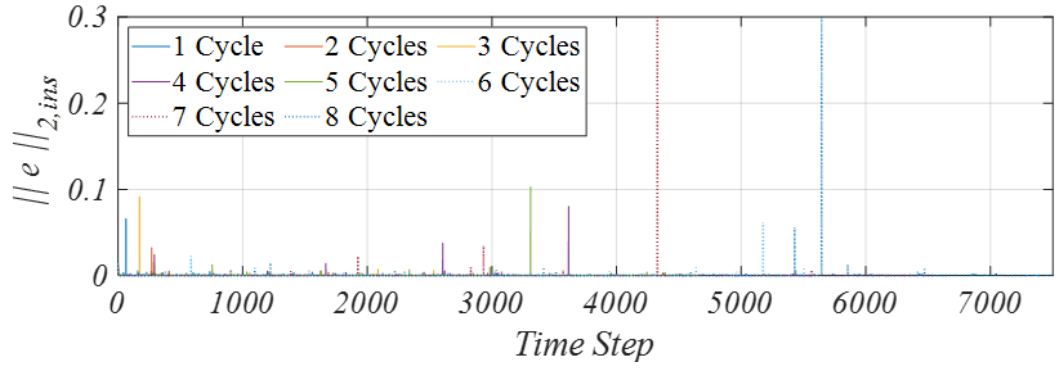

a)

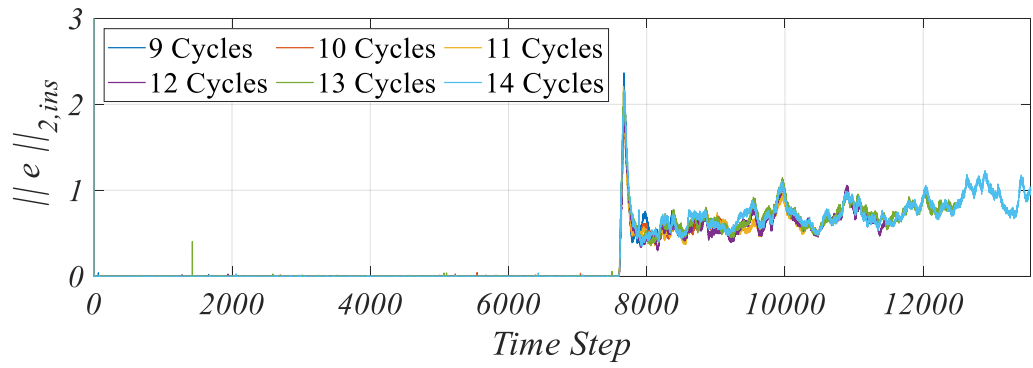

b)

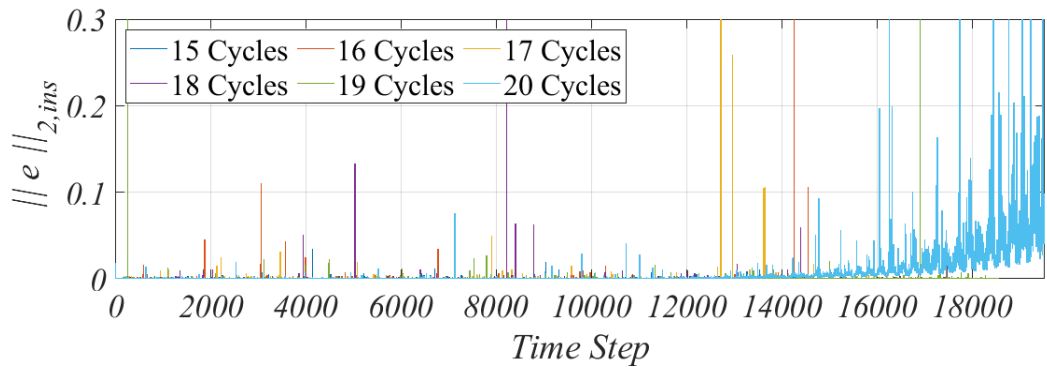

c)

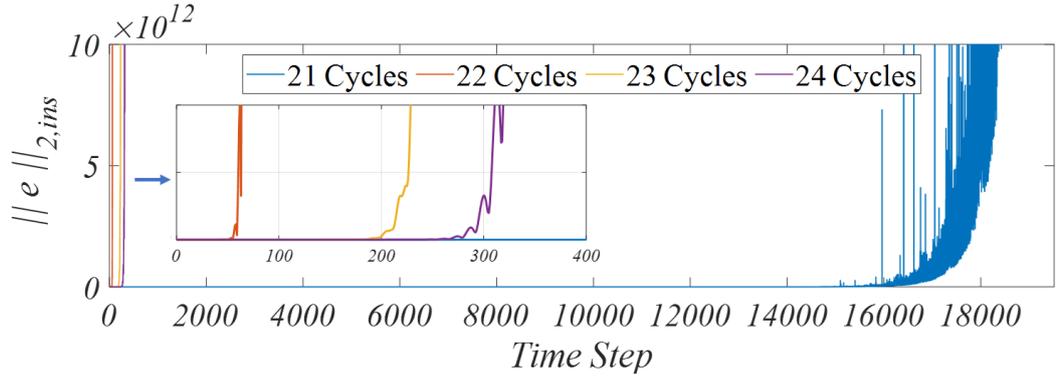

**Fig. 16** Mean $l_2$-norm of reconstruction error versus time step $t^*$ of a) 1-8 (Group 1), b) 9-14 (Group 2), c) 15-20 (Group 3), and d) 21-24 (Group 4) DMD-sampled oscillation cycles.

On the other hand, **Fig.** 16.b illustrates an entirely different trend. Conforming to the accuracy of Group 1 in the early stage, Group 2 universally exhibits a spike in error after *~7,800* time-steps. Afterward, the spike immediately drops off and plateaus on an elevated level *(~0.8).* In this range, not only is the accuracy slightly compromised, the spiking phenomenon *per se* is transfixing.

What causes the spike? First, we note a behavioral contradiction between *St* and reconstruction. Recall **Fig.** 13.b-c, the convergence of *St* for modes 2 and 3 begins after cycle 8, which coincides precisely with the timeline of the error spike. Before cycle 8, one would naturally assume that the fluctuations in *St* are due to the incomplete capture of flow mechanisms. However, the excellent reconstruction accuracy suggests the otherwise. To this end, we remark the sampling of a few cycles (Group 1) sufficiently capture all the flow dynamics, and the fluctuations in *St* are attributed to the DMD's pursuit of an optimal subspace. It is to say, with a small sample, although the DMD provides a sufficiently accurate description of the system, the set of descriptors (*i.e.,* DMD modes as Ritz pairs) are subjected to great variability. Sometimes, even the same set of descriptors are in play, it displays vastly different ranking in dominance. These observations imply the modal characteristics of the individual DMD mode are far from temporal convergence, and that $\tilde{A}$ is not invariant. Upon inspection of individual mode shapes, we confirmed this conclusion. So, Group 2 marks a transition stage in the algorithm's pursuit of optimal subspace.

Expectedly, Group 3 (**Fig.** 16.c) restores the reconstruction accuracy after the establishment of an optimal subspace. Up to Cycle 19, only occasional singularities violate the *0.1* threshold. This signifies the temporal convergence in both global dynamics and individual modal characteristics. $\tilde{A}$ is also invariant. By contrast, the curve of Cycle 20 begins to exhibit exponential growth after *16,000* time-steps, showing an early sign of divergence. Cycle 21 in Group 4 (**Fig.** 16.d) confirms this notion. The exponential growth becomes apparent and signifies a rapid deterioration of the DMD representation of the fluid system. It implies the Ritz pairs, manifested as the DMD modes, cease to provide good approximations of an invariant $\tilde{A}$.

One may also note cycles 22-24 exhibit the same exponential growth, except shooting off in the very early stage. We attribute the deterioration to the violation of the $m < n$ condition. Although cycle 21 sets a somewhat threshold of stability, the effect of divergence emerges as $m$ approaches $n$. Before the actual violation, the behavioral change is gradual, where the reconstruction accuracy in the early stage remains intact. Upon violation, the behavioral change is bi-polar, so the DMD representation becomes spurious and entirely meaningless.

### 4.3.2 The Convergence States

To better generalize our findings, we define the grand mean $l_2$-norm of reconstruction error from **Eq.** 4.3.1.1:

$$G_{\|e\|_2} = \frac{1}{m} \sum_{i=1}^{m} \|e\|_{2,ins,i}$$

*(4.3.2.1)*

The grand mean is a spatiotemporally averaged statistic that essentializes the behaviors of the reconstruction error into a single index. **Fig.** 17 presents the grand mean error versus the number of cycles. The generalization of the convergence is lucid. We found four distinct states:

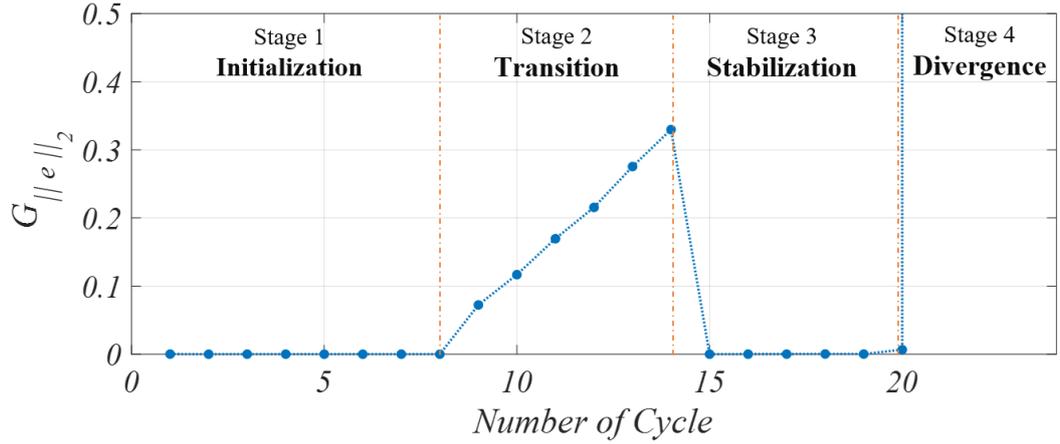

**Fig. 17** Grand mean $l_2$-norm of reconstruction error versus the number of DMD-sampled oscillation cycles.

Stage 1, the *Initialization*, defines the early stage in which, by sampling only a small range of data, the algorithm captures a system's spatiotemporal dynamics for fair reconstruction accuracy. However, the algorithm cannot find the optimal subspace nor a set of temporally stable Ritz descriptors, so the modal characteristics of individual modes are subject to great variability. The global $\tilde{A}$ has also yet to reach invariance. This stage is characterized by fluctuating $St$ and small reconstruction error.

Stage 2, the *Transition*, marks the intermediate phase in which the algorithm seeks the optimal subspace with more sampled data. Yet the tradeoff is a drastic deterioration of reconstruction accuracy. This stage is characterized by stabilizing $St$ and elevated reconstruction error.

Stage 3, the *Stabilization*, describes the optimal state in which the algorithm fully establishes the temporal convergence of modal characteristics and reconstruction accuracy. In this stage, sufficient data yields an optimal subspace and an invariant $\tilde{A}$, so the output of the decomposition becomes independent from additional sampling. This stage is characterized by near-constant $St$ and minimal reconstruction error.

Stage 4, the *Divergence*, depicts the case in which excessive sampling violates the DMD's tacit condition $m<n$, so the decomposition loses stability, and the algorithm yields meaningless output. This stage is characterized by the sudden loss of integrity in both the global and mode-specific DMD characteristics.

Finally, the benchmark reveals the sampling of 15-20 oscillation cycles achieves the *Stabilization* state for the turbulent flow herein, which appeals to a spatial-temporal ratio of *m~2/3n*. We also suggest the achievement of the *Stabilization* state for the most analytical effort on nonlinear systems, while the *Initialization* state suffices for most reconstruction tasks.

# 5. Parametrization of Range and Resolution

Based on the benchmark in previous section, we investigate the effect, both standalone from and combined with the sampling range, of the sampling resolution. We quantify the sampling resolution by a binary system $2^{SF}$ where $SF \in \mathbb{Z}^{0+}$. In practice, the temporal dimension $m$ dictates the upper limit of $SF$. The benchmark adopted the highest resolution, $SF=0$, which translates to the sampling of every single snapshot. **Table** 4 summarizes all the tested scenarios of $SF$.

**Table 4** The sampling resolution quantified by a binary system $2^{SF}$.

| $SF$ | Sampling Interval $(2^{SF})$ | Sampling Frequency $(St)$ |
|------|------|------|
| 0 | 1 | 124.5 |
| 1 | 2 | 62.22 |
| 2 | 4 | 31.11 |
| 3 | 8 | 15.56 |
| 4 | 16 | 7.778 |
| 5 | 32 | 3.889 |
| 6 | 64 | 1.945 |
| 7 | 128 | 0.9723 |
| 8 | 256 | 0.4862 |
| 9 | 512 | 0.2430 |
| 10 | 1024 | 0.1215 |
| 11 | 2048 | 0.0608 |

## 5.1 Convergence of Resolution

We first examine the standalone effect of the sampling resolution on the *Stabilization* state. **Fig**. 18 presents the Strouhal number *St* versus the sampling resolution of dominant modes 1-3 with 20 sampled cycles. The convergence of the sampling resolution is apparent up to *SF=8*. Close inspection reveals that *SF=6* is more appropriate because the constancy of *St* is slightly altered thereafter for modes 1 and 2. Given the global vortex-shedding frequency is *St=0.127*, mode 1 (*St=0.127*) and mode 2 (*St=0.121*) describe the Strouhal and Bloor-Gerrard instabilities that make up the Kármán vortex, respectively [64]. *SF=6* resolves the oscillation periods of modes 1 and 2 by using approximately 15 frames per cycle. We extend this observation to a suggestion of 15 frames per cycle for most fluid systems. Additionally, we advise users to tailor the resolution according to the dynamics of interest. For example, if one investigates the Kelvin-Helmholtz instability, *SF=6* (*St=1.945*) might be insufficient given its high-frequency, so the DMD may produce unsatisfactory results [48], [65]. However, an even coarser *SF=7* (*St=0.9723*) is like to be sufficient to resolve the periodic motions of separation bubbles [66].

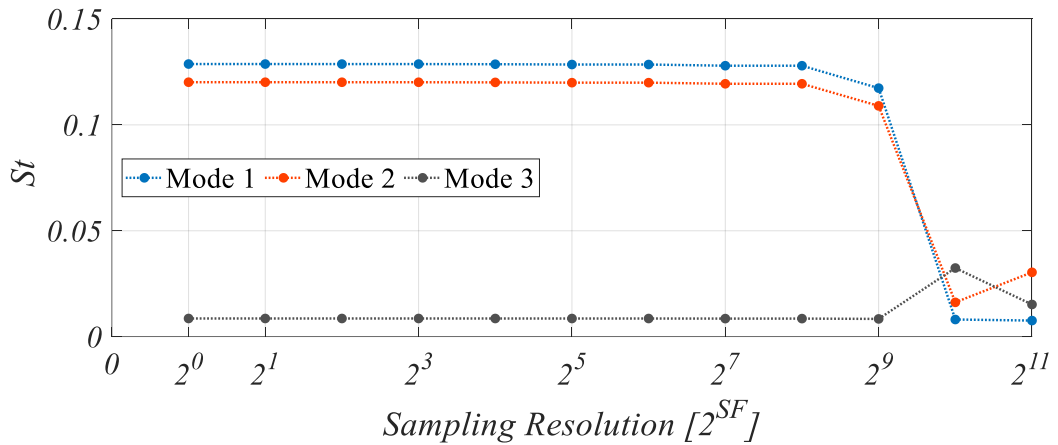

**Fig. 18** The Strouhal number versus the sampling resolution $2^{SF}$ of dominant Modes 1-3 for 20 DMD-sampled oscillation cycles.

## 5.2 Combined Effect of Range and Resolution

In this last section, we examine the combined effect of the sampling range and resolution by simultaneously parametrizing the number of cycles and $2^{SF}$. **Fig.** 19 presents the Strouhal number $St$ after omitting the unstable range (Stage 4). For all three modes, the observations are consistent with previous findings. Cycle 8 generally marks the beginning of the $St$ stabilization (*Transition* state), and $SF=6$ marks the end of the resolution convergence. More importantly, the figures show that, within the convergence state, variations in range and/or resolution do not affect the DMD output. This is to say, there is no apparent interaction or nonlinear complications between the sampling range and resolution. However, one shall reckon that the convergence of both is necessary for an optimal DMD decomposition.

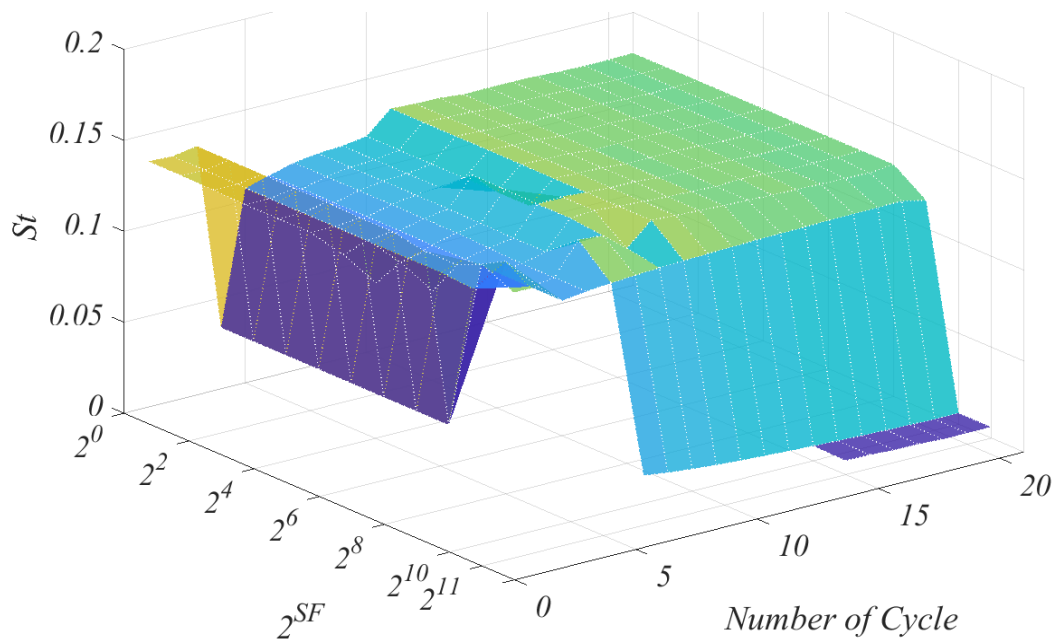

a)

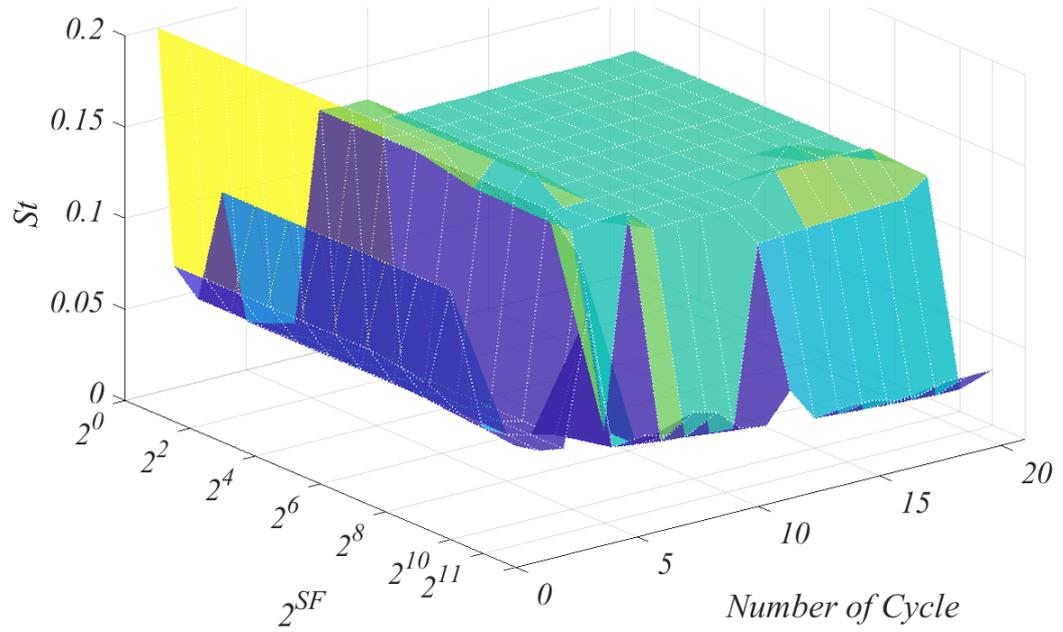

b)

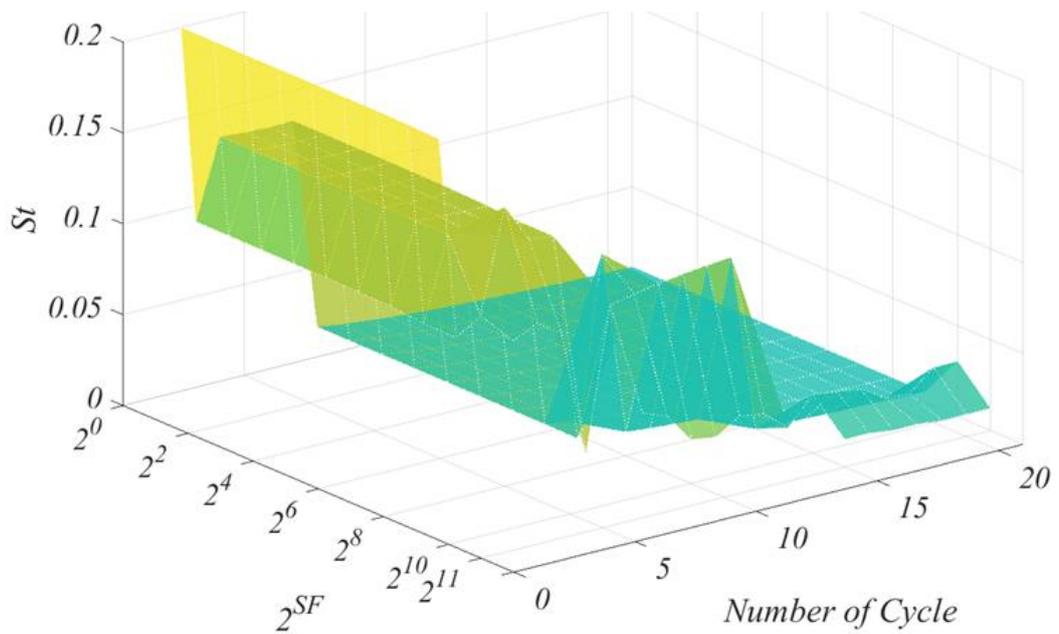

c)

**Fig. 19** The Strouhal number versus the sampling resolution $2^{SF}$ versus the number of DMD-sampled oscillation cycles of dominant a) Mode 1, b) Mode 2, and c) Mode 3.

After all, we draw several key remarks from this bi-parametric study:

- The convergence of the sampling range depends primarily on the global state of the system. The modal behaviors shift universally as the system transitions across the four states.

- The convergence of the sampling resolution depends primarily on the mode-specific periodicity. The convergence of one mode does not necessarily translate to that of another.

- For the fluid system herein, the convergence of the sampling range and resolution are mutually independent.

- For most analytical engineering implementations of the DMD, users are advised to sample a sufficient range to reach the *Stabilization* state without violating the $m<n$ condition and resolve the periodicity of target dynamics by at least 15 frames.

- For most reconstruction tasks, the *Initialization* state suffices but the convergence of sampling resolution shall be maintained. The *Transition* and *Instability* state shall generally be avoided.

## Conclusion

In this effort, we parametrically investigated the nuances of DMD sampling from an engineering point-of-view. We presented a parametric work, and, from which some practical suggestions on the sampling range and resolution for the DMD algorithm are provided. Based on high-fidelity LES-NWR data of a canonical fluid system, we found the convergence domains for both the sampling range and resolution, on which the DMD output is independent and $\tilde{A}$ is invariant to the respective variables.

Specifically, we discovered four states of mode convergence: *Initialization, Transition, Stabilization,* and *Divergence* depending on the sampling range. The *Stabilization* is the preferred state in which the decomposition reaches full statistical and modal convergence. Meanwhile, the *Initialization* suffices for most system reconstruction tasks. In addition, contrary to popular beliefs, excessive sampling violates the $m<n$ condition and produces meaningless DMD output. Finally, the convergence of the sampling range is a global transition of the system.

On the other hand, the convergence of the sampling resolution is mode-specific—the convergence of one mode does not necessarily translate to those of the others. We found an appropriate practical rule-of-thumb, which suggests that an oscillatory cycle of a target dynamical activity shall be resolved by 15 frames. Besides, we observed no apparent entanglement between the sampling range and resolution.

Finally, we must clarify that the conclusions drawn based on the model herein may not be exactly the same for its experimental counterpart, especially given the disparities in the restrictions of numerical and experimental procedures. A direction of future work can study the sensitivity of the parameters through empirical data to assess the applicability of the conclusions herein.

## Acknowledgements

We give a special thanks to the IT Office of the Department of Civil and Environmental Engineering at the Hong Kong University of Science and Technology. Its support for the installation, testing, and maintenance of our high-performance servers are indispensable for the current project. The work described in this paper was supported by a grant from the Research Grants Council of the Hong Kong Special Administrative Region, China (Project No. 16207719).

# Appendix I: Large-Eddy Simulation with Near-Wall Resolution

## AI.1 Large-Eddy Simulation

### AI.1.1 Governing equations

We simulated the turbulent flows by the Large-Eddy Simulation with Near-Wall Resolution. The LES-NWR does not employ any modeling in the near-wall viscous regions, hence the high accuracy. It is, however, computationally prohibitive for many high-*Re* flows. But for the moderate-*Re* subcritical regime herein, LES-NWR is feasible. In formulation, the filtered, time-dependent Navier-Stokes equations for an incompressible Newtonian fluid resolve the large-scale eddies and govern the fluid dynamics, which include the continuity equation:

$$\frac{\partial \bar{u}_i}{\partial x_i} = 0$$

*(AI.1.1.1)*

and the momentum equation:

$$\frac{\partial \bar{u}_i}{\partial t} + \frac{\partial \bar{u}_i \bar{u}_j}{\partial x_j} = -\frac{1}{\rho}\frac{\partial \bar{p}}{\partial x_j} + \nu \frac{\partial^2 \bar{u}_j}{\partial x_i \partial x_i} - \frac{\partial \tau_{ij}^{sgs}}{\partial x_i}$$

*(AI.1.1.2)*

where $\bar{u}_i$ denotes the filtered velocities; $\rho$ denotes the fluid density; $\nu$ denotes the fluid molecular viscosity; $\bar{p}$ denotes the modified pressure; $\tau_{ij}^{sgs}$ denotes the deviatoric subgrid stress.

The filtered equations differ slightly from the original Navier-Stokes, insofar as the continuity equation commutes, but an additional pseudo-stress term arises in the momentum equation. Like the Reynolds stress, the subgrid stress $\tau_{ij}^{SGS}$ originates from the nonlinear convection term. One may also decompose it into the isotropic and deviatoric parts:

$$\tau_{ij}^{SGS} \equiv \frac{1}{3}\tau_{kk}^{SGS}\delta_{ij} + \tau_{ij}^{sgs}$$

*(AI.1.1.3)*

While the former straightforwardly merges into the modified pressure term $\bar{p}$, the latter instigated an entire sub-domain of mathematical modeling called the subgrid stress models.

## AI.1.2 Smagorinsky-Lilly Model

Perhaps the most widely accepted subgrid stress model, as adopted in this work, is the Smagorinsky model [67] with the Lilly formulation [68]. The Smagorinsky model consists of two parts. First, the linear eddy-viscosity model by the Boussinesq hypothesis relates the deviatoric subgrid stress to the filtered rate of strain $\bar{S}_{ij}$:

$$\tau_{ij}^{sgs} = -2\nu_{sgs}\,\bar{S}_{ij} \qquad\qquad (A.1.2.1)$$

where the coefficient $\nu_{sgs}$ is named the subgrid viscosity. Henceforth, one may model the subgrid viscosity by the mixing-length hypothesis:

$$\nu_{sgs} = l_s^{\,2}\,\bar{S} \qquad\qquad (A.1.2.2)$$

where

$$l_s = min(\kappa d, C_s\Delta) \qquad\qquad (A.1.2.3)$$

denotes the Smagorinsky lengthscale; $\bar{S}$ denotes the filtered characteristic rate of strain; $\kappa = 0.40$ denotes the von Kármán constant; $d$ denotes the distance from a cell centroid to the closest wall; $C_s$ denotes the Smagorinsky constant; $\Delta$ denotes the filter width. For filtering, we adopted a grid-dependent filter in the physical space, $\Delta = V^{1/3}$, where $V$ denotes the cell volume. Also, $l_s$ takes in the smaller of $d$ and $\Delta$ to accommodate the near-wall viscous regions.

## AI.1.3 Dynamic-Stress Model

The original Lilly formulation derived a universal Smagorinsky constant $C_s = 0.17$, yet experiments found *'excessive damping of large scale fluctuations in the presence of mean shear and transitional flows as near-solid boundary'* [69]. Despite efforts to find the optimal value, one shall reckon that $C_s$ appeals to no universality and depends on the scale of fluid motions. Decades later, Germano *et al.* [70] and Lilly [1] developed a dynamic concept for $C_s$, which we have adopted herein. The dynamic-stress model imposes a second filter $\tilde{\Delta} = 2\Delta$ to compute the Smagorinsky constant based on the dynamics of the inter-filter eddies. The grid- and test-filtered stresses are:

$$\tau_{ij}^{SGS} = \overline{u_i u_j} - \overline{u_i}\,\overline{u_j}$$

*(AI.1.3.1)*

$$\mathcal{T}_{ij}^{SGS} = \widetilde{\overline{u_i u_j}} - \widetilde{\overline{u_i}}\,\widetilde{\overline{u_j}}$$

*(AI.1.3.2)*

One may also derive the deviatoric part as:

$$\tau_{ij}^{sgs} = -2\mathcal{C}_s \Delta^2 \overline{S}\,\overline{S}_{ij}$$

*(AI.1.3.3)*

$$\mathcal{T}_{ij}^{sgs} = -2\mathcal{C}_s \widetilde{\Delta}^2 \widetilde{\overline{S}}\,\widetilde{\overline{S}}_{ij}$$

*(AI.1.3.4)*

where $\mathcal{C}_s$ replaces $C_s$ in the original equations and may take in negative values to accommodate backscatter – the energy transfer from the subgrid scale to the resolved scale in depiction of the inverse energy cascade.

The Germano identity [70] relates the stresses by:

$$\mathcal{L}_{ij} = \mathcal{T}_{ij}^{SGS} - \widetilde{\tau_{ij}^{SGS}} = \widetilde{\overline{u_i}\,\overline{u_j}} - \widetilde{\overline{u}}_i\,\widetilde{\overline{u}}_j$$

*(AI.1.3.5)*

Taking $\mathcal{C}_s$ as uniform, one may define:

$$M_{ij} \equiv 2\Delta^2 \widetilde{\overline{S}\,\overline{S}_{ij}} - 2\widetilde{\Delta}^2 \widetilde{\overline{S}}\,\widetilde{\overline{S}}_{ij}$$

*(AI.1.3.6)*

Substituting **Eq.** *AI*.1.3.4 into **Eq.** *AI*.1.3.5 and combining with **Eq.** *AI*.1.3.6, one obtains:

$$\mathcal{L}_{ij}^S \equiv \mathcal{T}_{ij}^{sgs} - \widetilde{\tau_{ij}^{sgs}} = \mathcal{C}_s M_{ij}$$

*(AI.1.3.7)*

which is the Smagorinsky model for the deviatoric part of $\mathcal{L}_{ij}$,

$$\mathcal{L}_{ij}^d \equiv \mathcal{L}_{ij} - \frac{1}{3}\mathcal{L}_{kk}\delta_{ij}$$

*(AI.1.3.8)*

At last, the dynamic coefficient $\mathcal{C}_s$ takes the minimized mean-square error that best approximates $\mathcal{L}_{ij}^d$ by $\mathcal{L}_{ij}^S$ [1]

$$\mathcal{C}_s = \frac{M_{ij} L_{ij}}{M_{kl} M_{kl}}$$

*(AI.1.3.9)*

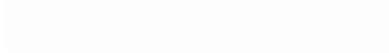

## Funding

The work described in this paper was supported by the Research Grants Council of the Hong Kong Special Administrative Region, China (Project No. 16207719).

## Conflict of Interest

The authors declare that they have no conflict of interest.

## Availability of Data and Material

The datasets generated during and/or analyzed during the current work are restricted by provisions of the funding source but are available from the corresponding author on reasonable request.

## Code Availability

The custom code used during and/or analyzed during the current work are restricted by provisions of the funding source.

## Author Contributions

All authors contributed to the study conception and design. Funding, project management, and supervision were performed by Tim K.T. Tse and Zengshun Chen. Material preparation, data collection, and formal analysis were led by Cruz Y. Li and Zengshun Chen, and assisted by Asiri Umenga Weerasuriya, Xuelin Zhang, Yunfei Fu, and Xisheng Lin. The first draft of the manuscript was written by Cruz Y. Li and all authors commented on previous versions of the manuscript. All authors read, contributed, and approved the final manuscript.

## Compliance with Ethical Standards

All procedures performed in this work were in accordance with the ethical standards of the institutional and/or national research committee and with the 1964 Helsinki declaration and its later amendments or comparable ethical standards.

## Consent to Participate

Informed consent was obtained from all individual participants included in the study.

## Consent for Publication

Publication consent was obtained from all individual participants included in the study.